\documentclass[aps,prl,onecolumn,groupedaddress,notitlepage,10pt]{revtex4-1}

\usepackage[T1]{fontenc}
\usepackage{graphicx}
\usepackage[labelfont=bf]{caption}
\usepackage{color}
\usepackage{hyperref}
\usepackage{amsmath}
\usepackage{amssymb}
\usepackage{setspace}

\newcommand{\bfr}{{\bf r}}
\newcommand{\bfk}{{\bf k}}

\newcommand{\ee}{\mathrm{e}}
\newcommand{\ii}{\mathrm{i}}

\begin{document}


\title{Phase-sensitive fluorescent imaging with coherent reconstruction}


\author{Jeffrey J.~Field}
\affiliation{W.M. Keck Laboratory for Raman Imaging of Cell-to-Cell Communications, Colorado State University, Fort Collins, CO 80523}
\affiliation{Department of Electrical and Computer Engineering, Colorado State University, Fort Collins, CO 80523 \\ \href{mailto:jeff.field@colostate.edu}{\color{blue} jeff.field@colostate.edu}}

\author{David G.~Winters}
\affiliation{W.M. Keck Laboratory for Raman Imaging of Cell-to-Cell Communications, Colorado State University, Fort Collins, CO 80523}
\affiliation{Department of Electrical and Computer Engineering, Colorado State University, Fort Collins, CO 80523}

\author{Randy A.~Bartels}
\affiliation{W.M. Keck Laboratory for Raman Imaging of Cell-to-Cell Communications, Colorado State University, Fort Collins, CO 80523}
\affiliation{Department of Electrical and Computer Engineering, Colorado State University, Fort Collins, CO 80523}
\affiliation{School of Biomedical Engineering, Colorado State University, Fort Collins, CO 80523}


\date{\today}

\begin{abstract}
Optical imaging plays a critical role in advancing our understanding of three dimensional dynamics of biological systems. Coherent imaging (CI) methods exploit spatial phase information, encoded through propagation of coherent signal light emerging from a specimen, to extract a three-dimensional representation of the object from a single high-speed measurement. Until now, CI methods could not be applied to incoherent light, severely limiting their ability to image the most powerful biological probes available -- fluorescent molecules -- with sufficient speed and volume to observe important processes, such as neural processing in live specimens. We introduce a new imaging technique that transfers the spatial propagation phase of coherent illumination light to incoherent fluorescent light emission. The transfer of propagation phase allows CI techniques to be applied to fluorescent light imaging, and leads to large increases in imaging speed and depth of field. With this advance, biological imaging of fluorescent molecules is significantly expanded.
\end{abstract}


\maketitle

\section*{Introduction}
The ability to spatially resolve the structure of objects plays a critical role in understanding nearly every physical system, providing insights into fields as diverse as astrophysics \cite{McKinney:2013uu,Moran:2004tb}, molecular solvation dynamics \cite{Dickson:1996wd}, geophysical systems \cite{Wolfe:2009ug}, the behavior of solid-state materials \cite{Chen:2013vg}, and biological systems \cite{Olivier:2010vk}. Likewise, the ability to resolve temporal dynamics is essential for unravelling processes of physical systems in relation to internal and external stimuli. Obtaining a comprehensive understanding of the interplay between spatial organization and functional behavior in dynamic specimens remains a pressing challenge, particularly in biological systems. 
	
Many biological processes are mediated chemically, and relevant mechanisms are often revealed only by labelling such processes with fluorescent probes \cite{Zhang:2002ws}. Biological sciences rely on the molecular specificity of fluorescent probes to study targeted chemical processes. Three dimensional images formed by collection of incoherent light can be obtained with high spatial resolution using techniques that restrict illumination and/or detection to a small volume in the object region \cite{Davidovits:1969tda,Denk:1990wsa}. However, restricting data collection to a small volume necessitates serial data acquisition, limiting imaging speed and degrading the ability to track dynamic behaviors in 3D. 
	
With notable exception \cite{Diebold:2013uya}, coherent radiation enables much faster 3D imaging than is currently possible with incoherent light. Three dimensional images of an object can be reconstructed from a single, rapidly acquired, two dimensional image of light collected from spatial points throughout the object by exploiting our understanding of coherent wave diffraction behavior. This powerful tool has proven useful in seismic, ultrasonic \cite{Devaney:1982vhb}, photo-acoustic \cite{Xu:2011wb}, optical \cite{Leith:1963wna,Seaberg:2011ttb}, x-ray \cite{Sandberg:2008vk,Miao:2012tbb}, and electron-beam \cite{Raines:2010vw} imaging systems for rapid detection of spatial organization and temporal dynamics. These techniques, generally referred to as coherent imaging (CI), rely on the recording, or numerical recovery, of optical spatial phase information that reveals the distance the light has travelled from the point at which it scatters in an object to where it is recorded.
	
Special techniques are required to extract the phase and amplitude of the diffracted signal field because optical detectors are incapable of responding to the phase of a light beam. Optical interference converts phase differences between two beams to intensity variations that can be recorded and used to recover the amplitude and phase. Holographic microscopy (HM) \cite{Leith:1963wna} represents the current state-of-the-art \cite{Smith:2013uq}. Alternatively, the amplitude and phase of a coherent field diffracted from an object can be recovered through inversion of coherent light propagation with a regularized, iterative optimization algorithm \cite{Sandberg:2008vk,Miao:2012tbb,Raines:2010vw}, without requiring interference.
	
Despite the advances of these techniques for imaging coherent radiation in a three dimensional region, the problem of dynamic volumetric imaging of 3D incoherent light emission, such as from fluorophores, remains vexing. The phase of the fluorescent light emitted by the molecules bears no stable relationship to the illumination light or neighboring fluorophores in the specimen, rendering the fluorescent light completely spatially incoherent. As a result, no phase information is available to pinpoint the location of the fluorescent emitter, normally excluding the application of CI techniques to fluorescent imaging. The vast potential of CI has lead to its application to holography using incoherent sources \cite{Lohmann:1965uza,Peters:1966uk,Cochran:1966upa,Bryngdahl:1968tab} and fluorescent emission \cite{Poon:1985wrb,Schilling:1997tnb,Rosen:2008tj}. Such techniques fail to image in the presence of optical scattering, with fluorescent light emission at high speed, or with large volumes in an epi-collection configuration, all of which are needed for monitoring dynamic biological behaviors. 

CI is able to image 3D volumes at high speeds, but these powerful tools have not been applicable to incoherent light imaging until now. In this Article, we introduce an imaging method that applies the high speed 3D imaging capabilities of CI to incoherent fluorescent light emission. Encoding the spatial phase of a coherent illumination source into a measurement of fluorescent light intensity collected with a single-element photodetector makes this possible. Dubbed Coherent Holographic Image Reconstruction by Phase Transfer (CHIRPT), this technique enables application of the full range of strategies used for coherent volume imaging with incoherent light. CHIRPT can, in principle, be applied to any incoherent emission or scattering process that is illuminated by a coherent source. CHIRPT imaging will prove particularly valuable for biological applications because of its ability to image at high acquisition rates in an epi-fluorescence configuration, its relative insensitivity to scattering due to single-element detection, and ability to image a large depth-of-field.

\section*{Physical principles of CHIRPT}
CHIRPT enables CI techniques to be used with incoherent light. The illuminating light is modulated with a temporal intensity pattern unique to each spatial point in the object region. The modulation is formed through interference of coherent illumination beams, and this interference transfers the spatial propagation phase of the illuminating fields to the temporal modulation pattern. The modulated illumination light excites fluorescent molecules, and the light emitted by those molecules mirrors the temporal illumination intensity pattern at the molecule location. The fluorescent light emitted by the object is collected onto a single element photodetector, and the signal generated by the photodetector contains information that identifies the position of fluorescent molecules in the object. The signal is a compilation of intensity modulation patterns emitted by fluorescent molecules in the object. With this set of modulation patterns, spatial propagation phase in the object can be recovered through a simple, direct computation. Using this spatial phase, we can now exploit the power of CI techniques to form images with incoherent light.

\begin{figure}[ht]
\begin{center}
\resizebox{4.5in}{!}{\includegraphics{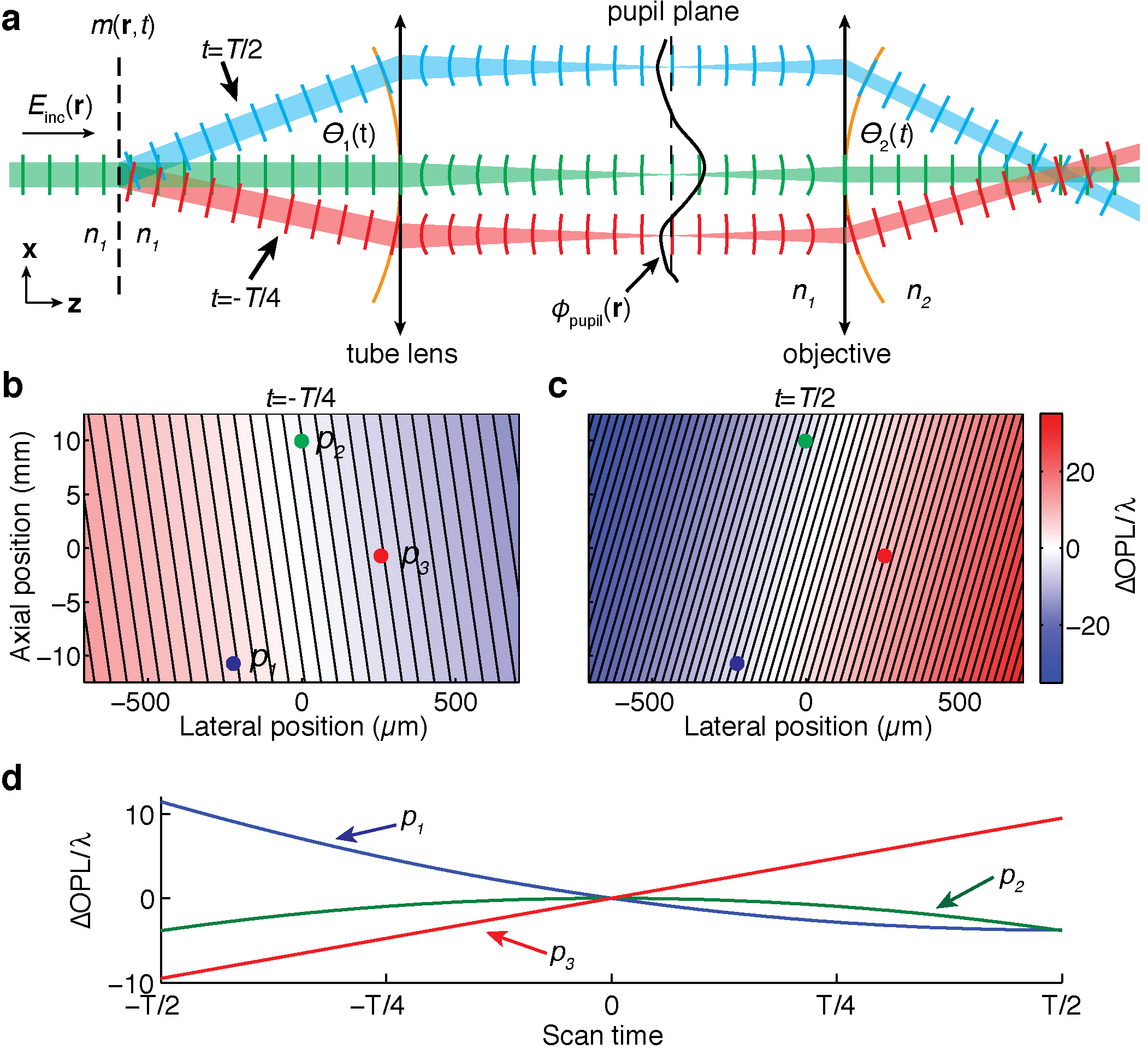}}
\caption{\label{fig:system-concept} Principle of CHIRPT imaging. (a) The angle of diffraction from the modulator, $\theta_1(t)$, varies with scan time with periodicity $T \equiv 1/\nu_r$. Scan times of $-T/4$ and $T/2$ are indicated by the red and blue beams, respectively. (b) The difference in optical path length, $\Delta \mathrm{OPL}/ \lambda$, between the diffracted and undiffracted beams at $t = -T/4$, and (c) $t = T/2$. Contour lines represent integer values of $\Delta \mathrm{OPL}/\lambda$, corresponding to peaks in the illumination intensity pattern. (d) $\Delta \mathrm{OPL}/\lambda$ as a function of scan time for the points $(p_1,p_2,p_3)$ plotted in (b) and (c). Here, $\Delta \mathrm{OPL}$ was computed assuming a paraxial imaging system, where $\Delta \Phi(x,z;t) \approx 2 \pi \, \Delta x \,  (M \, \kappa_1 \, t) - \pi \, \lambda \, \Delta z \, (M \, \kappa_1 \, t)^2$.}
\end{center}
\end{figure}

Generating the intensity modulation pattern is the key aspect of CHIRPT, which is shown conceptually in Fig.~\ref{fig:system-concept}. We begin by directing a laser beam into a cylindrical lens to form a line focus. The line focus passes through a modulator disk spun at a constant angular velocity. The modulator disk is lithographically printed with a mask that contains a transmission grating designed to vary the groove density (spatial frequency) with angular position (Fig.~S1). As the laser line passes through the spinning disk, a portion of the laser power diffracts to an angle, $\theta_1(t)$, determined by the groove density. The undiffracted zero order beam is undeviated upon propagation through the mask, and thus stationary through the duration of the scan. The two beams -- diffracted and undiffracted -- are imaged to the object region as light sheets whose crossing angle varies with time, as represented in Fig.~\ref{fig:system-concept}a. It is the interference between these beams that creates the intensity modulation pattern.
	
The evolution of the interference pattern in the object plane $(x,z)$ with time is dictated by the relative optical path, $\Delta \mathrm{OPL}(x,z;t)$, accumulated by the two beams as they propagate from the mask plane to the object region. We can understand the OPL acquired by each beam using the framework provided by Fermat's principle of least action \cite{Born:1999un}. The OPL of the undiffracted beam is fixed, whereas the diffracted beam OPL depends on the diffraction angle. An ideal imaging system introduces zero OPL difference for paths taken between conjugate points in the modulation and object planes. OPL variations caused by imperfect imaging are captured by the pupil aberration phase, $\phi_\mathrm{pupil}(r)$. This aberration phase causes further variation of the illumination intensity, which can be numerically corrected with CHIRPT (Fig.~S4).
	
Along with the imaging system OPL, the beams pick up additional OPL as they diffract along the optic axis ($z$) away from the object image plane. The contribution of the imaging system OPL and the diffraction OPL determines the total OPL difference, $\Delta \mathrm{OPL}$, between the two beams. A derivation of the illumination intensity modulations, $I_\mathrm{ill}(x,z;t)$, caused by $\Delta \mathrm{OPL}$ shows the intensity follows the form $I_\mathrm{ill}(x,z;t) \propto \cos[\Delta \Phi(x,z;t)]$. Here, the phase difference is a function of $\Delta \mathrm{OPL}$ and the wavelength of the illumination light, $\lambda$, given by $\Delta \Phi(x,z;t) = 2 \pi \, \Delta \mathrm{OPL}(x,z;t)/\lambda$.
	
With an understanding of the contributions to the OPL described above, we can now write out a simple model of the CHIRPT imaging process. The relative phase is given by (Supplementary Information):
\begin{equation}
\Delta \Phi(x,z;t) = \Delta \Phi_a(t) +  2 \pi \, \Delta x \, \left(\frac{n_2}{n_1} M \, \kappa_1 \, t \right)  + \, 2 \pi \, \, \frac{n_2}{\lambda} \, \Delta z \left[ \sqrt{ 1 - \left( \frac{\lambda}{n_1} \, M \, \kappa_1 \, t \right)^2 } - 1 \right] 
\label{eq:phase_diff}
\end{equation}
where $\Delta \Phi_a(t)$ is the difference in aberration phase between the undiffracted and scanned beams as a function of time, and $n_2$ and $n_1$ are the refractive indices in the object region and the mask region respectively. Here $\Delta x$ is the lateral location relative to the centroid of the illumination intensity distribution; $\kappa_1 \equiv \Delta k \, \nu_r$ is the chirp parameter introduced by the modulation mask, describing the linear change in temporal modulation frequency as a function of lateral coordinate; $\Delta k$ is a parameter of the mask, setting the highest spatial frequency of the pattern; $\nu_r$ is the rotational frequency of the disk; $M$ is the magnification of the image relay optics, defined as the ratio tube lens focal length to the objective lens focal length; and $\Delta z$ is the axial location relative to the focal plane of the imaging system.

The illumination intensity is composed of a temporally-varying component, $I_1(x,z;t)$, and a stationary component, $I_0$ (Supplementary Information). Only the time dependent portion of the illumination intensity forms CHIRPT images. Retaining just the time dependent portion leads to an illumination intensity of:
\begin{equation}
I_1(x,z;t) = \frac{1}{\pi} \, \cos \left\{\Delta \Phi_a(t) + 2 \pi \, \Delta x \, \left(\frac{n_2}{n_1} M \, \kappa_1 \, t \right)  + \, 2 \pi \, \frac{n_2}{\lambda} \, \Delta z \, \left[\sqrt{1 -\left(\frac{\lambda}{n_1} \, M \, \kappa_1 \, t \right)^2} - 1 \right]  \right\}
\label{eq:intensity}
\end{equation}

Equation (\ref{eq:intensity}) describes how each position $(x,z)$ in the object plane experiences a unique illumination intensity modulation pattern that varies with scan time. At each scan time, the intensity variations show uniform, tilted fringes representing a single spatial frequency in the illumination intensity. The uniform fringe spacing is caused by a uniform variation in $\Delta \mathrm{OPL}$. Figure~\ref{fig:system-concept}b and Fig.~\ref{fig:system-concept}c demonstrate this for the two scan times indicated in Fig.~\ref{fig:system-concept}a, where $\Delta \mathrm{OPL}$ was computed from the theoretical analysis. Contour lines on the OPL maps in Fig.~\ref{fig:system-concept}b and Fig.~\ref{fig:system-concept}c correspond to the peaks of the illumination intensity pattern in the object region.

The OPL variation evolves with time as the disk spins. Each point in the object region experiences a unique temporal variation in $\Delta \mathrm{OPL}$. A visual representation of this can be seen for the three points indicated in Fig.~\ref{fig:system-concept}b and Fig.~\ref{fig:system-concept}c, where the full temporal trace gives a variation used to uniquely identify each spatial location. These OPL variations drive the intensity modulations at each point. Light emitted by fluorescent molecules excited at each point follows the illumination intensity temporal variation, provided the modulation frequencies are below the inverse fluorescent lifetime (Supplementary Information).

\begin{figure}[ht]
\begin{center}
\resizebox{4.95in}{!}{\includegraphics{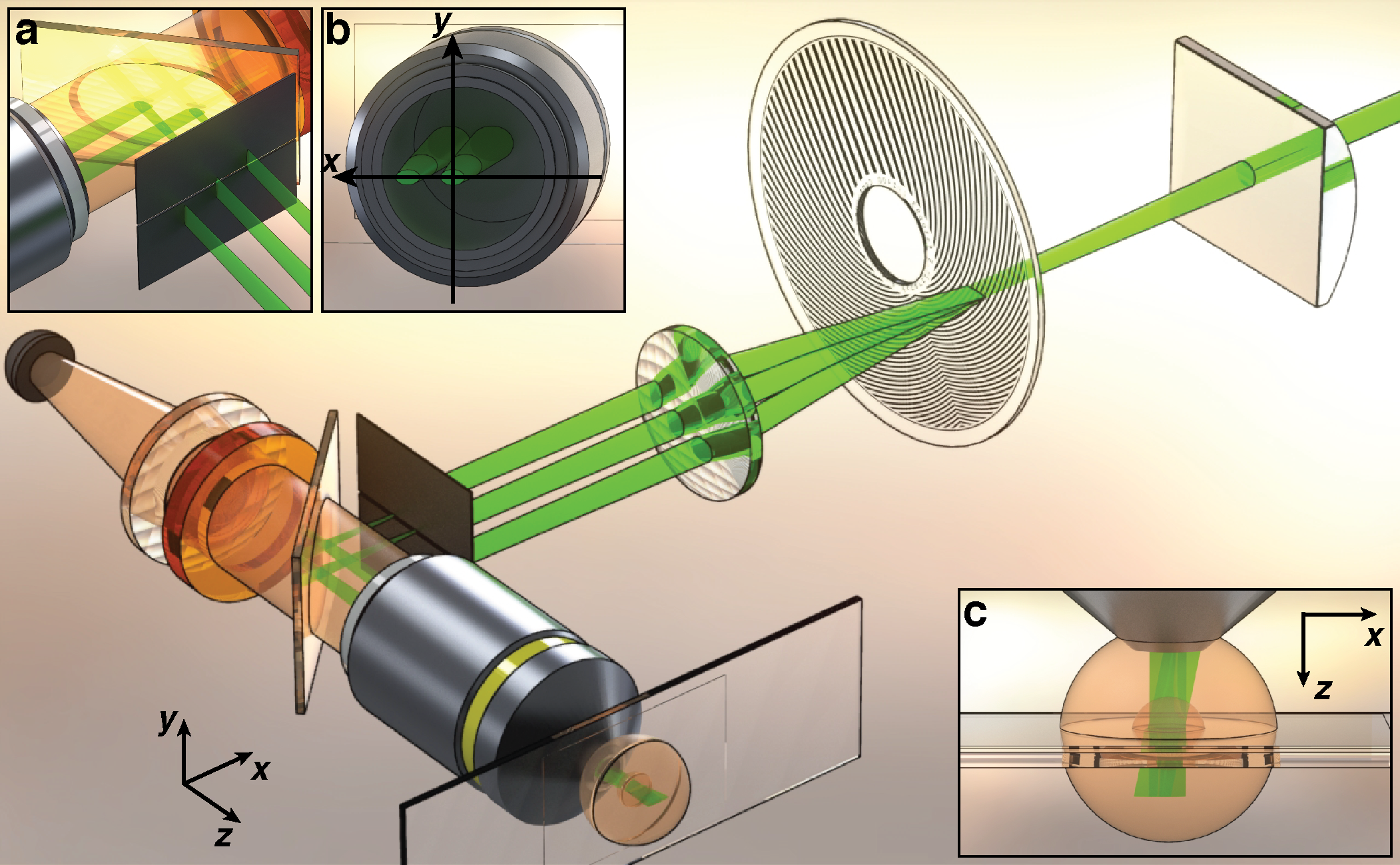}}
\caption{\label{fig:schematic} Schematic of the CHIRPT setup. (a) A constant diffraction angle in the vertical dimension results in vertical offset of the diffracted beam from the undiffracted beam (Supplementary Information). A horizontal slit is placed near the pupil plane of the objective lens to select a portion of the undiffracted beam and one diffracted order. (b) The illumination objective is oriented such that the illumination beams propagate along $y = 0$. (c) The diffracted and undiffracted beams interfere in the object region to generate a unique modulation intensity pattern for each point in the $(x,z)$ plane.}
\end{center}
\end{figure}

Most of the energy diffracted from the mask is contained in the undeviated zero-order beam ($j = 0$) and the first diffracted orders ($j = \pm1$). If the pupil phase is symmetric about the optic axis, e.g., spherical aberration, the two diffracted orders carry redundant and time-reversed spatial phase information. This redundancy in phase information introduces a spatial phase sign ambiguity akin to the twin-image problem that arises in on-axis holographic imaging. Unambiguous recovery of the spatial phase of the illumination system requires retaining only one diffracted order in CHIRPT imaging. We accomplish this by obstructing one of the first diffracted orders, in practice the $j = -1$ beam, in a Fourier plane of the modulation mask (Fig.~\ref{fig:schematic}). A portion of the $j = 0$ and $j = +1$ beams are spatially filtered in a plane conjugate to the back focal plane of the objective lens. The modulation mask facilitates spatial filtering of one diffracted order because it imparts a constant, non-zero spatial frequency in the vertical dimension, allowing complete spatial separation of the positive and negative diffracted orders in the filtering plane (Supplementary Information).

CHIRPT signals are formed by collecting all of the object fluorescent light on a single element detector that integrates the emitted object intensity. The total signal is a superposition of temporal intensity modulation patterns with an amplitude determined by the local fluorophore concentration, $C(x,z)$, and is expressed mathematically as: $S(t) = \int \, \mathrm{d}z \, \mathrm{d}x \, \, I_{\mathrm{ill}}(x,z;t) \, C(x,z)$ . Assuming the emitted fluorescence intensity has the same temporal structure as the illumination pattern, the instantaneous frequency of the signal recorded from the photodetector encodes the location of a fluorescent emitter in the $(x,z)$ plane. The instantaneous frequency for a single emitter is: 
\begin{eqnarray}
\nu(x,z;t) & = & \frac{\partial}{\partial t} \left\{\frac{\Delta \Phi(x,z;t)}{2 \pi} \right\} = \frac{\partial}{\partial t} \left\{\frac{\Delta \mathrm{OPL}(x,z;t)}{\lambda} \right\} \nonumber \\
	& = & \Delta x \, \left(\frac{n_2}{n_1} \, M \, \kappa_1 \right) - \frac{n_2}{n_1^2} \frac{\lambda \, \Delta z \, \left( M \, \kappa_1 \right)^2 \, t}{\sqrt{1 - \left(\lambda \, M \, \kappa_1 \, t/n_1 \right)^2}}  + \frac{1}{2 \pi} \frac{\mathrm{d} \Delta \Phi_a(t)}{\mathrm{d}t}
\label{eq:freq}
\end{eqnarray}

The average instantaneous frequency of the illumination modulation encodes the transverse emitter position, $\Delta x$, while deviations in the instantaneous frequency over the scan time provide a unique determination of the axial position, $\Delta z$.

For an object composed of many emitters, the temporal patterns from each emitter in the object add together, and they must be separated to recover the object spatial information, $C(x,z)$. Fluorescent emitters that lie in the object image plane, $\Delta z = 0$, exhibit a uniform modulation frequency that depends on the lateral position, $\Delta x$. The temporal pattern for emitters axially displaced from the object plane carries an additional temporal phase. In CHIRPT, the temporal domain is proportional to lateral spatial frequency, so the temporal CHIRPT signal represents a superposition of frequency-domain Fresnel zone plates (FZP) in the paraxial optics limit. By noting that holography forms images through superposition of spatial FZPs \cite{Rogers:1950wf}, we can use an algorithm similar to that used for holographic imaging to numerically construct CHIRPT images (Supplementary Information). The CHIRPT algorithm selects the positive frequency sideband of the signal, $S(t)$, then applies a propagation spatial frequency phase to recover the object, $C(x,z)$.

CHIRPT transfers the phase evolution of coherent illumination beams propagating through the object region onto a temporal pattern of fluorescent light emission. The spatial phase information can be recovered by numerically processing the CHIRPT signal, allowing CI techniques to be directly applied to incoherent light. The CHIRPT signals can be numerically refocused to reconstruct object information in the entire $(x,z)$ plane. The phase information also directly records imaging system aberration phase that can be numerically removed, eliminating optical imaging aberration distortions from CHIRPT images (Fig.~S4). CHIRPT signals simultaneously collect fluorescent emission from a large range of depths in a single measurement, allowing the collection of out of focus contributions that would not be captured with a conventional imaging system, as well as digital refocusing of blurry objects. The result is that CHIRPT images have a large depth of field.

\section*{Results}
For a detailed description of the experimental methods, please consult the Supplementary Information. Briefly, we used a continuous-wave laser with a 532~nm wavelength to illuminate various fluorescent objects. Modulation masks with density $\Delta k = 70/\mathrm{mm}$ were used for all images shown in this work. The magnification and numerical aperture of the imaging system varied with the chosen objective lens. In the data presented here we used either a 50$\times$/0.8~NA or a 20$\times$/0.45~NA objective lens, resulting in an overall system magnification of 95$\times$ and 30.4$\times$ respectively. Both objectives were air-immersion, so in the data presented here we set $n_1 = n_2 = 1$. Fluorescent light was measured in the epi-direction with a photomultiplier tube (PMT). The signal from the PMT was electronically amplified and filtered prior to digitization with a data acquisition (DAQ) card. All components of the microscope and data collection were controlled with a custom C\# application written in house. Please consult Table.~S1 for a summary of data acquisition times for each image presented here. 

\begin{figure}[ht]
\begin{center}
\resizebox{\linewidth}{!}{\includegraphics{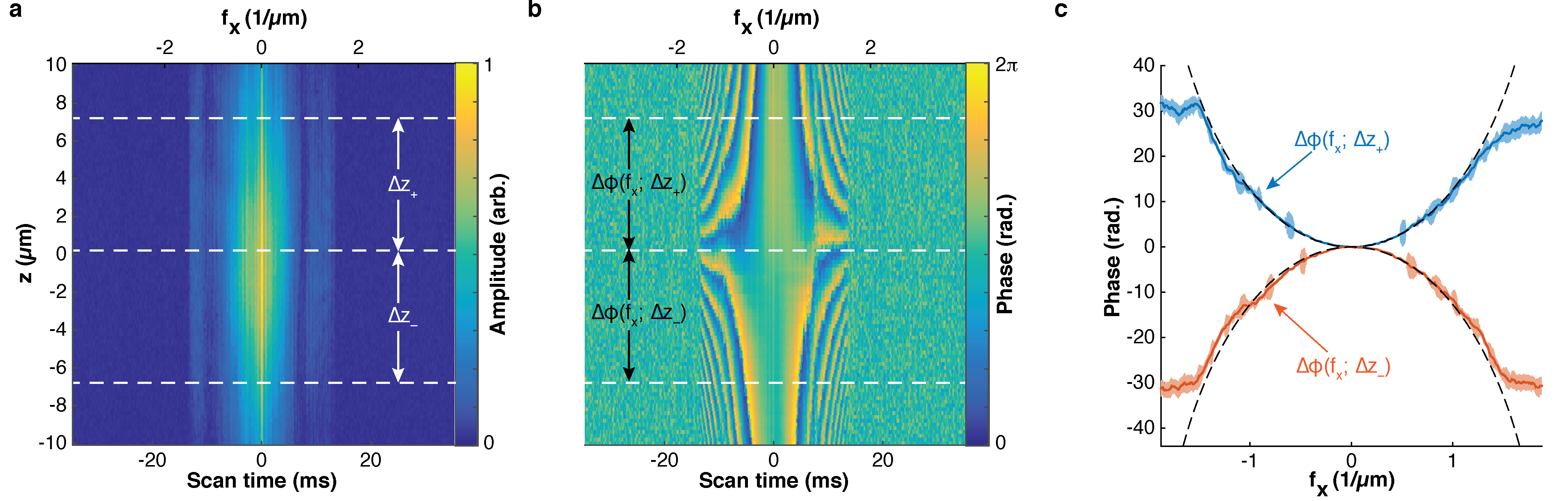}}
\caption{\label{fig:phase-data} The spatial phase difference between the illumination beams is encoded in the CHIRPT signal. (a) Amplitude and (b) phase of the complex CHIRPT data from a fluorescent nanodiamond as a function of physical defocus. (c) Positive and negative defocus phase with respect to the phase measured at the nominal focus (blue and red solid lines respectively). Shaded areas denote the standard deviation obtained by measuring 20 consecutive scans. Dashed lines correspond to the theoretical Ewald phase for the defocus values read from the axial translation stage encoder. Data was collected at a scan rate of 17~frames/s. The total acquisition time for all 20 images was 119~s.}
\end{center}
\end{figure}

To validate the CHIRPT theory, we imaged a single 100-nm-diameter fluorescent nanodiamond (FND) in the $(x,z)$ plane. The fluorescence intensity distribution through the focus was measured by axially translating the FND along the direction of light propagation and recording the CHIRPT signal at each location. The complex CHIRPT signal was obtained from the measured voltage by a Hilbert transform (Fig.~S3). The amplitude and phase of the complex CHIRPT signal are shown in Fig.~\ref{fig:phase-data}a and Fig.~\ref{fig:phase-data}b respectively. Equation~\ref{eq:phase_diff} predicts that the phase evolution as a function of defocus should follow the Ewald circle shifted by the axial phase of the undiffracted beam. To validate this prediction, we examined the phase difference for two defocus positions relative to the nominal focal plane, indicated by dashed lines in Fig.~\ref{fig:phase-data}a and Fig.~\ref{fig:phase-data}b. The phase differences for the positive and negative defocus values are shown in Fig.~\ref{fig:phase-data}c with solid lines, while the theoretical phase difference predicted by Eq.~\ref{eq:phase_diff} is plotted in dashed lines. The strong agreement between the measured and theoretical phase differences verified that the defocus phase was encoded in the temporal measurement as predicted by the theory. 

\begin{figure}[ht]
\begin{center}
\resizebox{5.75in}{!}{\includegraphics{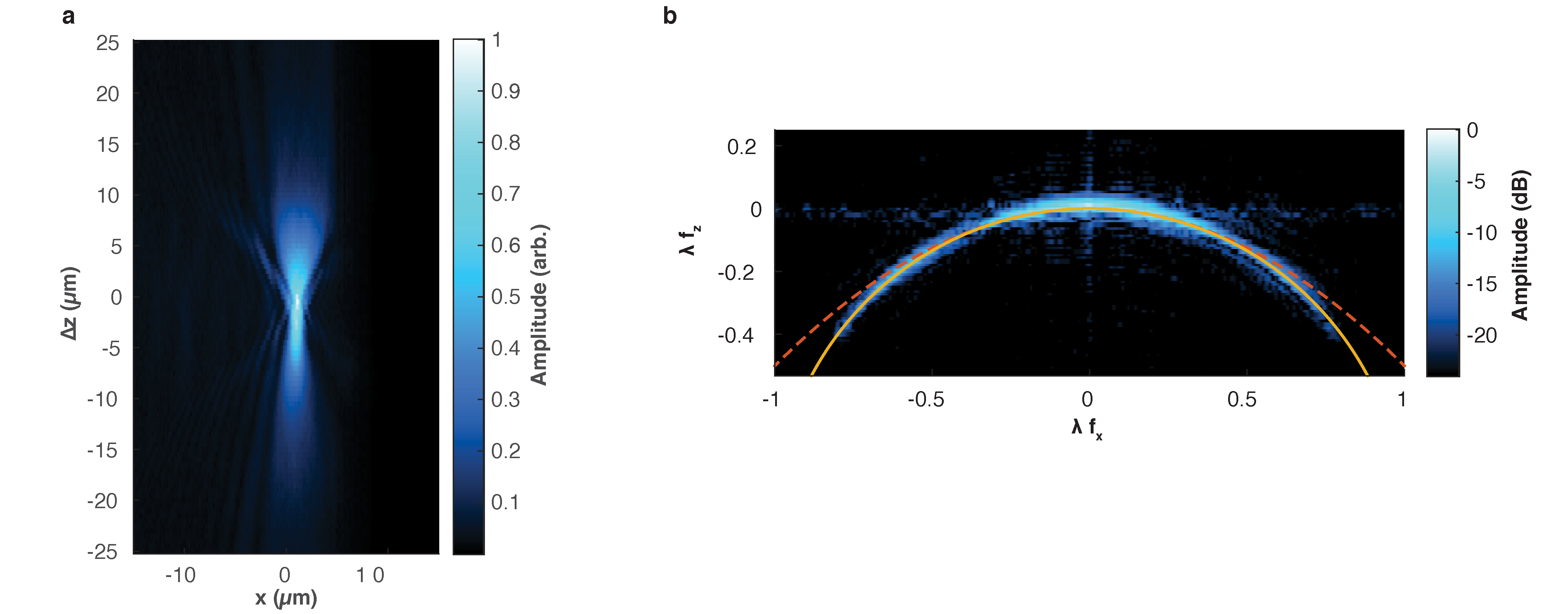}}
\caption{\label{fig:ewald} CHIRPT encodes the Ewald phase as a function of physical defocus. (a) Fluorescence intensity vs. physical defocus measured by translating a 100 nm FND along the axial dimension. Data was collected with a 95$\times$/0.8 NA CHIRPT system. This image is the average of 25 images with a total collection time of 148.9 s. (b) A two-dimensional FFT of the image in (a) reveals the axial phase encoded as a function of defocus. The phase closely follows the Ewald phase (solid line). The Fresnel phase (dashed line) is also displayed to show that the paraxial approximation breaks down under these imaging conditions, so data must be numerically propagated with the Ewald phase. }
\end{center}
\end{figure}

Measuring the intensity of a sub-diffraction-limited fluorescent object as a function of system defocus gives the point-spread function (PSF) and the corresponding optical transfer function (OTF) for our CHIRPT microscope in the $(x,z)$ plane. Both the PSF and OTF are displayed in Fig.~\ref{fig:ewald}. The curvature of the OTF is further confirmation that the axial phase evolution in the CHIRPT microscope is well described by the Ewald phase.

\begin{figure}[ht]
\begin{center}
\resizebox{3.5in}{!}{\includegraphics{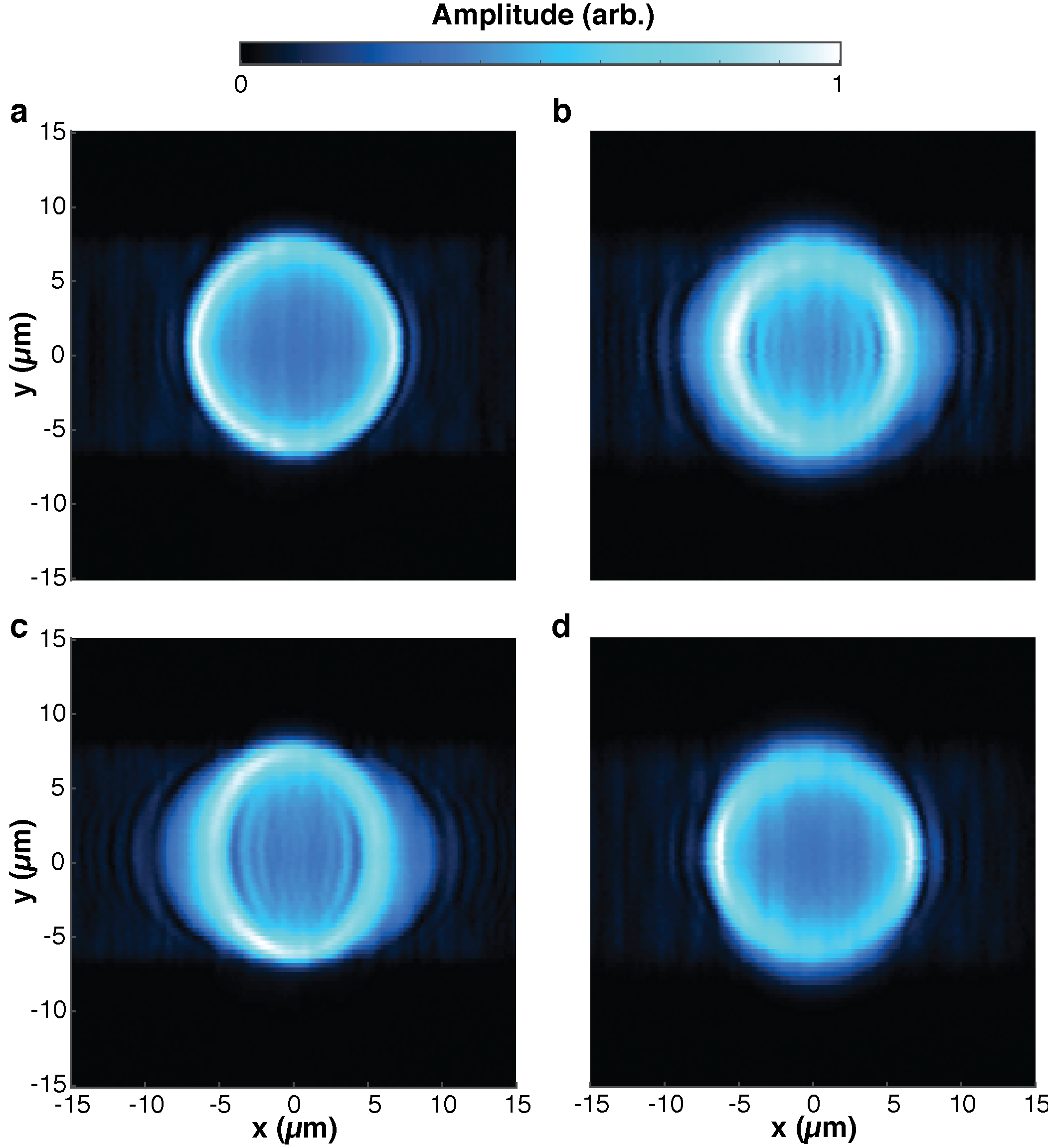}}
\caption{\label{fig:bead-refocus} Computed reconstruction of a fluorescent bead. (a) Fluorescence image of an in-focus 15~$\mu$m shell-stained fluorescent bead. (b) Image of the same bead with -10.17~$\mu$m of physical defocus (towards the objective lens). (c) Image formed by numerically propagating the image in (a) by -10.17~$\mu$m in the axial dimension, and similarly, (d) the image formed by numerically propagating the image in (b) by +10.17~$\mu$m. Images in (a) and (b) are averages of 15 images, acquired at a frame rate of 17~frames/s with total acquisition time of 89.3~s. }
\end{center}
\end{figure}

To further confirm the spatial phase measurement from fluorescence intensity, we numerically reconstructed in-focus and out-of-focus 2D images in the $(x,y)$ plane of a single shell-stained fluorescent microsphere (Fig.~\ref{fig:bead-refocus}). The measured images can be propagated along the defocus distance by applying the conjugated Ewald phase, which is proportional to defocus, $\Delta z$ (Supplementary Information). The fluorescent sphere was first imaged in focus (Fig.~\ref{fig:bead-refocus}a), followed by an image collected after translating the bead in the defocus direction by -10.17 $\mu$m, i.e., toward the objective lens. Images shown in Fig.~\ref{fig:bead-refocus}c and Fig.~\ref{fig:bead-refocus}d were computed by refocusing the collected images in Fig.~\ref{fig:bead-refocus}a and Fig.~\ref{fig:bead-refocus}b by -10.17~$\mu$m and +10.17~$\mu$m respectively. Ideally, Fig~\ref{fig:bead-refocus}a and Fig.~\ref{fig:bead-refocus}d would be identical, as would Fig.~\ref{fig:bead-refocus}b and Fig.~\ref{fig:bead-refocus}c. The excellent agreement between these images is further confirmation that the physical defocus phase is indeed encoded into the measured CHIRPT signal. Higher-order spatial phase distortions imparted by the imaging system were also recovered from other CHIRPT data and used to perform digital aberration correction (Fig.~S4). 

\begin{figure}[ht]
\begin{center}
\resizebox{3.5in}{!}{\includegraphics{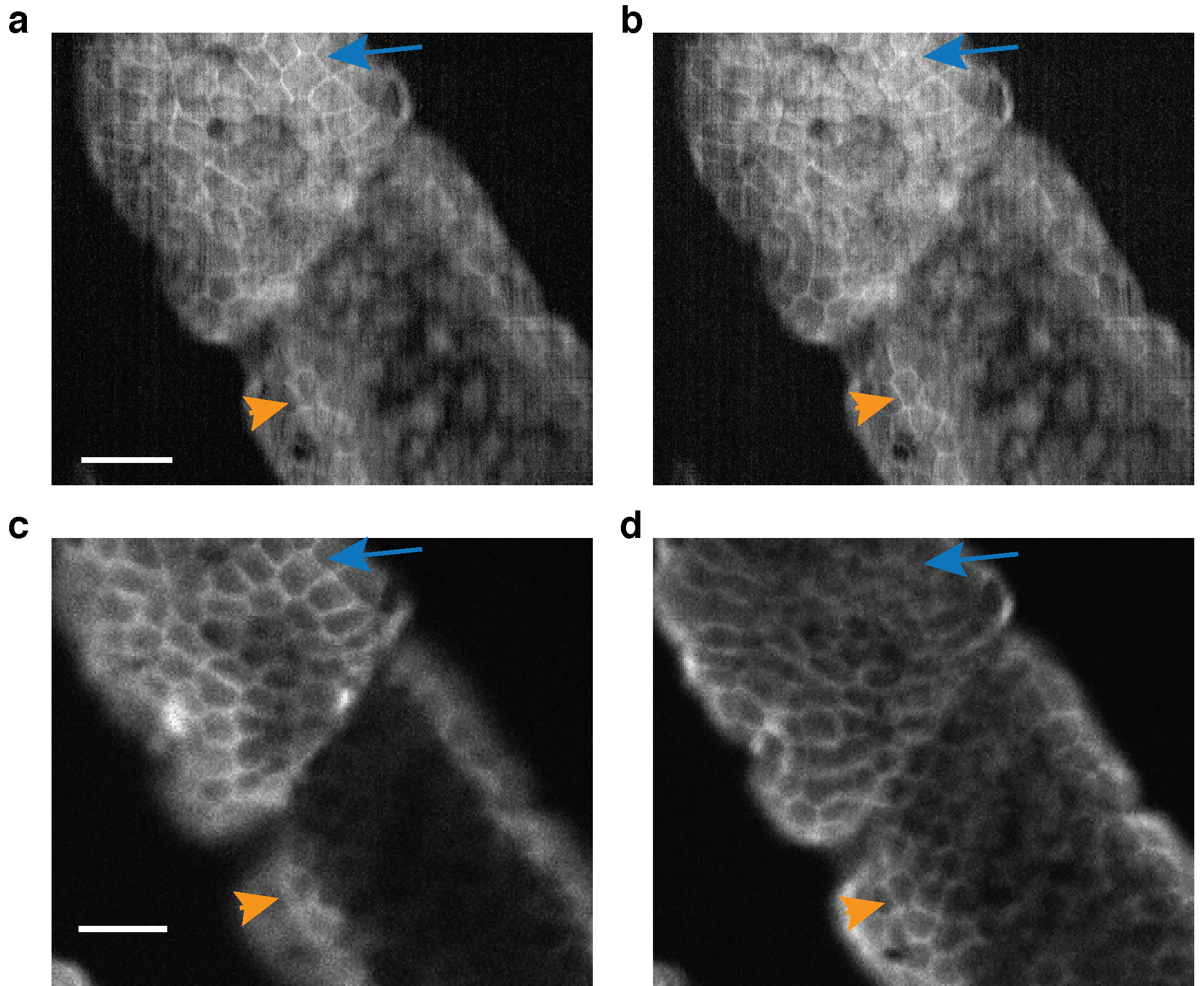}}
\caption{\label{fig:mouse-intestine} CHIRPT and confocal imaging of fluorescently-labelled murine intestine slices. (a) and (b) Digitally refocused CHIRPT images captured with 532~nm illumination light. Images were formed by propagating the average of 25 measured images. Panel (a) displays the image propagated by -4.5~$\mu$m while panel (b) displays the measured image. (c) and (d) Confocal images with 561~nm illumination light. Images were collected with an axial separation of 4.5~$\mu$m. In all four images, features that are in focus are denoted by arrows. Scale bars: 20~$\mu$m. CHIRPT data was acquired at a rate of 17~frames/s for a total acquisition time of 443.7~s.}
\end{center}
\end{figure}

To explore the ability of CHIRPT to refocus fluorescent images in biological specimens, epi-fluorescent images from 16-$\mu$m-thick fixed murine intestinal tissue were collected. Figure~\ref{fig:mouse-intestine}a shows a 2D image reconstruction formed by digitally refocusing a measured CHIRPT image by -4.5~$\mu$m, while Fig.~\ref{fig:mouse-intestine}b shows the measured image. Both images show features, marked by arrows, being brought into focus at two different depths in the tissue by digital propagation of fluorescence intensity. To benchmark CHIRPT imaging in the tissue slices, the same region was imaged with a commercial spinning-disk confocal imaging system (Fig.~\ref{fig:mouse-intestine}c and Fig.~\ref{fig:mouse-intestine}d). The CHIRPT images clearly show features in focus at different axial planes, consistent with the confocal images.

Comparison of the images in Fig.~\ref{fig:mouse-intestine} reveals differences in the nature of image formation between CHIRPT and confocal microscopy. While a single CHIRPT image reveals cells from a range of depths in the tissue, a confocal image contains data from only a single axial plane due to optical sectioning. This numeric refocusing ability gives CHIRPT a very large depth of field (DOF). We computed the DOF in CHIRPT imaging using geometric considerations, and found it to vary inversely with the numeric aperture of the imaging system, i.e., $\mathrm{DOF} \propto \mathrm{NA}^{-1}$ (Supplementary Information). Thus CHIRPT is able to image the location of fluorophores in the object of a much larger axial range than is possible with conventional imaging, for which $\mathrm{DOF} \propto \mathrm{NA}^{-2}$. 

\begin{figure}[ht]
\begin{center}
\resizebox{3.5in}{!}{\includegraphics{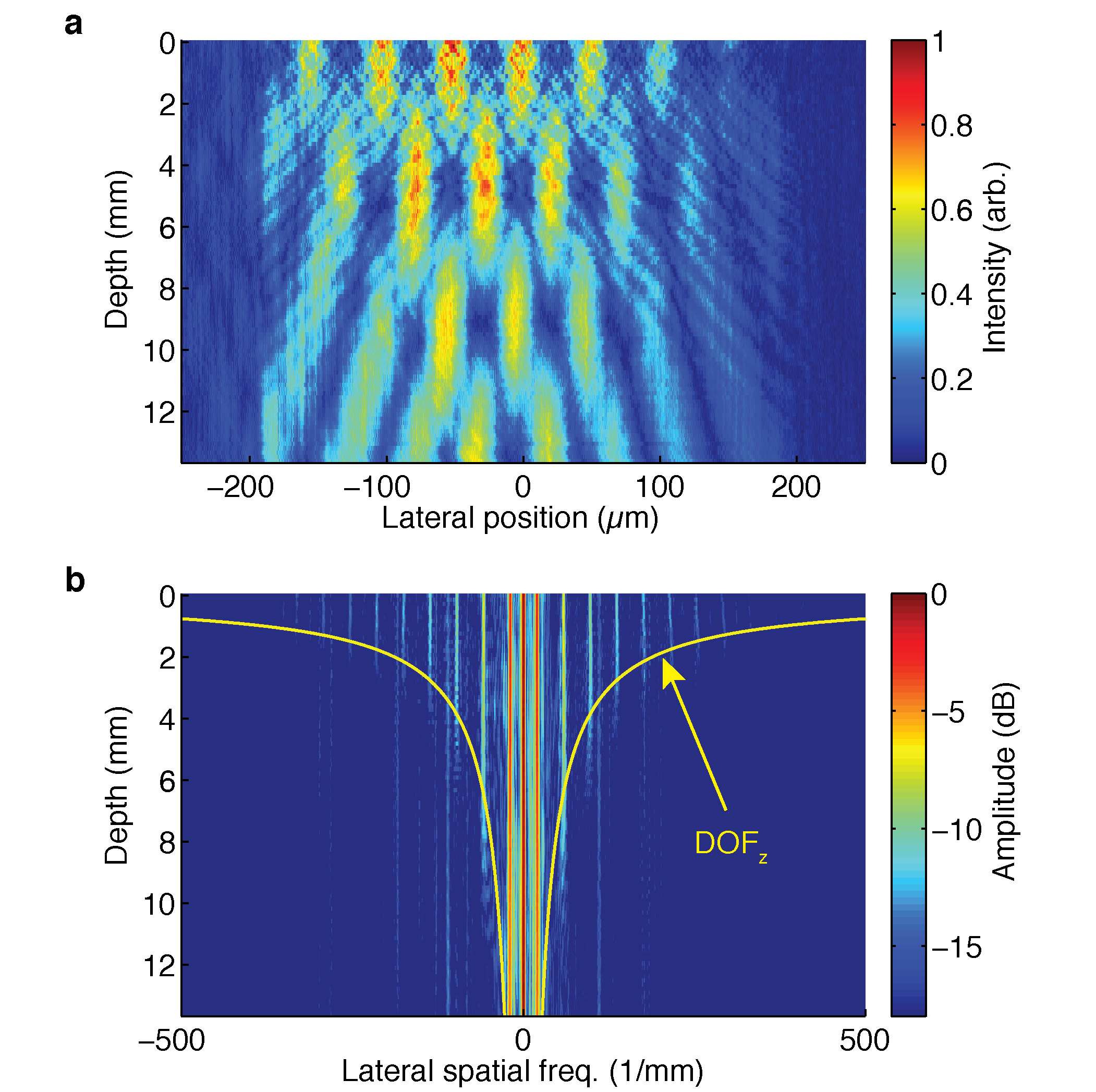}}
\caption{\label{fig:dof} Large depth of field (DOF) imaging of a 20 line/mm Ronchi ruling with CHIRPT. (a) Transmissive image as a function of depth, where a depth of 0 mm nominally corresponds to physical contact of the ruling with the illumination objective lens. (b) The 1D FFT of the image in (a) with respect to $x$, showing the decay of harmonics of the Ronchi ruling's fundamental spatial frequency with defocus. Data was measured at 31 frames/s, for a total image acquisition time of 3.23 s for the single-shot image shown in (a).}
\end{center}
\end{figure}

We measured the DOF in CHIRPT by collecting an image of the illumination laser transmitted through a 20 line/mm Ronchi ruling as the ruling was defocused (Fig.~\ref{fig:dof}). A 0.45~NA objective lens was used to resolve the object. The working distance of the objective lens is 510~$\mu$m, yet images were reliably formed though a scan range of 14~mm, or >27 times the working distance of the lens. Note that far away from the focal plane of the lens there is a loss of lateral spatial frequency content, made clear in Fig.~\ref{fig:dof}b by the odd harmonics of the binary, 50\% duty cycle Ronchi ruling have a depth-dependent amplitude. The yellow line indicates the on-axis depth of field as a function of lateral spatial frequency, $\mathrm{DOF}_z = 2 \, w/(\lambda \, f_x)$. The beam size, $w$, was estimated from the image in panel Fig.~\ref{fig:dof}a to be 100~$\mu$m. Since the largest spatial frequency passed by the system is related to the numerical aperture by $\mathrm{NA} = \lambda \, f_{x,\mathrm{max}}$, the on-axis DOF for the full resolution of the lens can be recast as: $\mathrm{DOF}_z = 2 \,  w/ \mathrm{NA}$.

We calculated the axial depth over which the full lateral spatial frequency support of the imaging system was retained to be approximately 440~$\mu$m, in good agreement with the data in Fig.~\ref{fig:dof}. This is approximately 83$\times$ larger than the predicted value of 5.3~$\mu$m for a conventional imaging system, defined as $\mathrm{DOF} \approx 2 \, n \, \lambda / \mathrm{NA}^2$ \cite{Muller:2005kq}.

\begin{figure}[ht]
\begin{center}
\resizebox{4.5in}{!}{\includegraphics{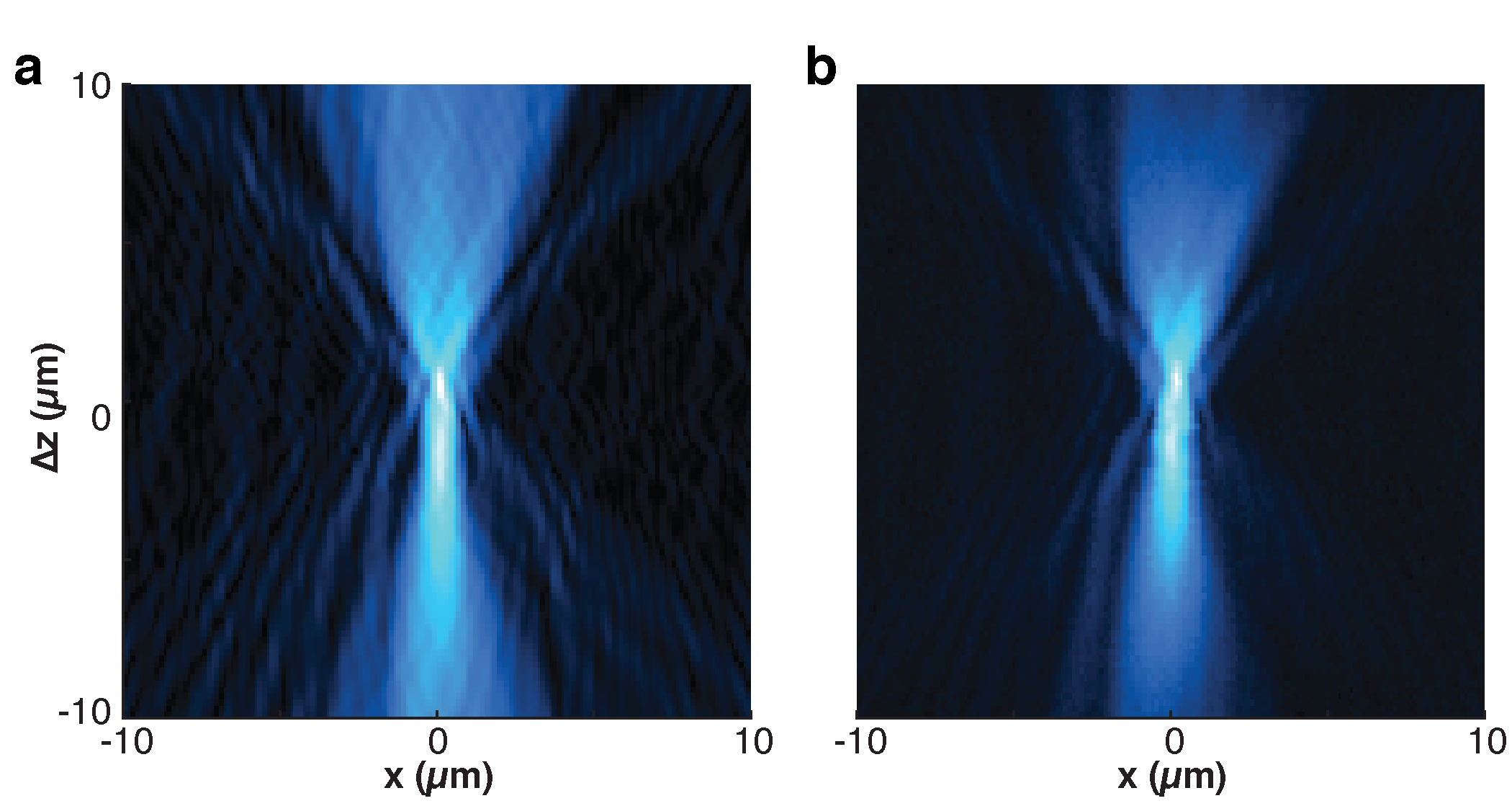}}
\caption{\label{fig:speed-comparison} Single-shot vs. averaged image acquisition of fluorescence form an FND. (a) A 2D image reconstructed from a single measured 1D image at 14.3~Hz. (b) An image of the same region in the $(x,z)$ plane formed by physically defocusing the FND and averaging twenty 1D images at each defocus location. Total acquisition times for (a) and (b) were 0.07~s and 141.4~s respectively.}
\end{center}
\end{figure}

For the majority of data presented here, we chose to display averaged images to improve the signal-to-noise ratio (SNR). This also helped to verify the theoretical analysis of the CHIRPT microscope, i.e., Fig.~\ref{fig:phase-data} and Fig.~\ref{fig:ewald}.  We note however that CHIRPT reliably formed images with single-shot data acquisition. An example of this can be found in Fig.~\ref{fig:speed-comparison}. While Fig.~\ref{fig:speed-comparison}a shows a 2D image in the $(x,z)$ plane reconstructed from data collected in a single rotation of the modulation mask, Fig.~\ref{fig:speed-comparison}b shows the average of 20 measured 1D images. The single shot image was collected in 0.07~s while the averaged image was acquired in 141.4~s, not including time added by stop-and-hold stage scanning. While the SNR of Fig.~\ref{fig:speed-comparison}b is indeed better than the single-shot image in Fig.~\ref{fig:speed-comparison}a, the single-shot image was acquired over 2,000 times faster. The fidelity of the reconstructed image is remarkable given the relatively fast frame rate of the acquired data. The SNR is the primary factor limiting image acquisition speed in all forms of biological microscopy, and CHIRPT is no exception. However, CHIRPT shows great promise for drastically improving 2D frame rates. Another example of high-speed single-shot imaging with CHIRPT is shown in Fig.~S6, where we imaged a fluorescent object with an acquisition time of 2.48~ms for a 2D $(x,z)$ plane. The full 3D image volume was acquired in approximately one second, ignoring the hold time of the linear positioning stages. 

Finally, we note that the spatial resolution of CHIRPT imaging in the $x$ dimension is diffraction limited. The maximal lateral spatial resolution of an imaging system, $\delta x$, is limited by the largest lateral spatial frequency that can be passed by the lens, $f_{x,\mathrm{max}}$. The NA of the objective lens is directly proportional to the maximal lateral spatial frequency that the lens can accommodate. To achieve the maximal resolution afforded by the objective lens in an imaging configuration, $f_{x,\mathrm{max}}$ must contribute to image formation, which occurs when the back aperture of the objective lens is filled. Since CHIRPT images are formed by illumination with varying lateral spatial frequencies as a function of scan time, which are generated by the spinning modulation mask, optimal lateral spatial resolution is achieved when the highest spatial frequency generated by the modulation mask and image relay optics is greater than $f_{x,\mathrm{max}}$. In that case, $\delta x$ is determined by the NA of the imaging objective and the wavelength of the illumination light, $\lambda$. Using the Abbe criterion, the resolution limit is $\delta x = \lambda/ (2 \, \mathrm{NA})$.

Conversely, we have constructed CHIRPT imaging systems for which the back-aperture of the objective lens is not filled. In that case the effective NA of the imaging system is limited by the largest spatial frequency generated by the modulation mask and the magnification of the imaging system. The effective NA of the system is related to the magnitude of the maximal lateral spatial frequency in the object region via: $\mathrm{NA} = \lambda f_{x,\mathrm{max}} = n_2 \, \sin \theta_2(\pm T/2)$, and the lateral resolution of the system is: 
\begin{equation}
\delta x = \frac{n_1}{n_2}\frac{1}{M \, \Delta k}.
\end{equation}

For all data presented in the manuscript, the back aperture of the objective lens was filled -- allowing imaging with diffraction-limited spatial resolution determined by the NA of the objective lens. The optimal lateral spatial resolution of each imaging system was computed using the Abbe criteria, results in a computed spatial resolution limit of 591~nm for the 20$\times$/0.45 NA lens, and 333 nm for the 50$\times$/0.8 NA lens. Note that while these resolution numbers correspond to uniform lateral spatial frequency support, the practical limit of the spatial resolution in CHIRPT is the amplitude of the measured signal as a function of scan time (spatial frequency). 

In the y direction, the spatial resolution is set by the effective NA determined by the filling of the pupil aperture along the y direction in the objective pupil plane. The y-resolution limit is also $\delta y = \lambda / (2 \,  \mathrm{NA})$.

\section*{Discussion}
CHIRPT introduces a fundamentally new approach to imaging: the transfer of coherent propagation phase of illuminating beams to incoherent light emission. This phase transfer enables use of a powerful set of CI tools that rely on spatial phase information to rapidly and efficiently record object information. These methods were previously not applicable to incoherent light imaging, which encompasses many of the most powerful tools in biological imaging, such as fluorescence and Raman scattering. We have demonstrated this new method and its ability to capture coherent phase information with recorded incoherent light using holographic imaging, which relies critically on phase information.
	
The approach in CHIRPT can be applied to a broad range of imaging modalities that use waves to form images. CHIRPT is highly achromatic, and can be applied to imaging radiation ranging from THz waves to x-rays. CHIRPT could prove particularly valuable for mid infrared (MIR) and terahertz (THz) imaging, where there is a dearth of imaging detectors. Moreover, the fundamental principles can be adapted beyond electromagnetic radiation, including acoustic waves used in seismic and ultrasound imaging, or in the hybrid acoustic/optical imaging technique photoacoustic microscopy \cite{Xu:2011wb}.

CHIRPT has particular value for imaging in biological specimens. We have demonstrated high speed imaging of fluorescent and absorptive objects, with the ability to collect fluorescent images in a backscattered (epi) direction. Epi-fluorescent collection in CHIRPT has the potential to greatly impact biology studies, as it allows for light-sheet-microscopy-like imaging \cite{Huisken:2009qf,Chen:2014cr} with a single lens. Furthermore, the image is acquired with a single-pixel detector, which allows for robust imaging in the presence of optical scattering, such as with biological tissues. The single pixel detector, in combination with high-speed modulation allows for transformative improvements in imaging speed. We have shown $(x,z)$ plane collection in an epi-fluorescent configuration at 400~planes/second (Fig.~S6), and have achieved $>$600~planes/second with multiplexed modulation masks (data not shown). CHIRPT allows collection of highly defocused light, enabling imaging volumes much larger than conventional approaches. An exceptional depth of field, approaching 30 times the working distance of the objective lens was demonstrated -- well beyond that capable in conventional fluorescent imaging methods, particularly for 3D imaging. This set of capabilities brings new possibilities for multi-scale spatial and temporal imaging of biological organisms.

Application of this technology to high speed imaging of biological dynamics and to large-scale imaging of biological specimens, including large tissue regions and ultimately organs, will provide a new window that could help form a comprehensive understanding of biological behaviors. Not only does this advancement enable the powerful toolbox of CI techniques to be applied to imaging with incoherent light for the first time; it also enables the practical capabilities described above that permit new abilities in imaging.

\section*{Funding Information}
Funding was provided by the W.M.~Keck Foundation.

\section*{Acknowledgments}
The authors thank Henry Kapteyn, Margaret Murnane, Jeff A. Squier and Omid Masihzadeh for a critical reading of a version of the manuscript. The authors also thank James R. Bamburg for use of the confocal imaging system. JJF thanks Barbara W. Bernstein for assistance with confocal image acquisition. The fluorescent nanodiamond slurry was donated by Susanta Sarkar (Department of Physics, Colorado School of Mines, Golden CO).

\section*{Supplementary Information}

\setcounter{figure}{0}
\makeatletter
\makeatletter \renewcommand{\fnum@figure}
{\figurename~S\thefigure}
\makeatother

\section*{Experimental Methods}
\subsection*{CHIRPT Microscope}
Two different continuous wave diode-pumped solid state (DPSS) lasers were used in these experiments, both of which operate at a nominal wavelength of 532~nm.  Data in Fig.~7 and Fig.~S\ref{fig:fast_scanning} were acquired with a 100~mW source (Coherent Inc., Compass 315M-100). All other data was collected with a Sprout 5G pump laser (Lighthouse Photonics). 
	
Each laser was spatially filtered with a 10$\times$/0.25~NA objective lens (Zeiss A-Plan, UIS) and a 10-$\mu$m pinhole (ThorLabs, P10S). The output of the spatial filter was collimated with an 80-mm focal length achromatic lens (ThorLabs, ACH254-080-A-ML), yielding a beam with a 4.95~mm full-width, measured as the width between the 20\% and 80\% peak intensity points. The remainder of the microscope was different for each laser source.
	
For data collected with the Compass laser, the spatially-filtered beam was focused with a 300-mm focal length cylindrical lens, creating a line focus that extended laterally across the modulation mask (Fig.~\ref{fig:mask}), with a vertical height (full-width at the 20\%-80\% measurement points) of ~15~$\mu$m. After passing through the modulation mask, the modulated line focused beam was re-imaged and demagnified with a 250-mm tube lens (ThorLabs ACH-254-250-A-ML), and an objective lens. The objective lens used for collecting experimental data depended on the particular experiment. Data in Fig.~7 was collected with a 20$\times$/0.45~NA Zeiss A-Plan UIS objective lens, resulting in a system magnification of 30.4$\times$. High speed data presented in Fig.~S\ref{fig:fast_scanning} was collected with a 10$\times$/0.25~NA Zeiss A-Plan UIS objective lens, providing a system magnification of 15.2$\times$.
	
To increase the NA of the CHIRPT imaging system, the microscope was redesigned and rebuilt to allow for more precise control and alignment of the imaging system. In the redesigned CHIRPT system, we used the Lighthouse Photonics laser, as it has a relative intensity noise (RIN) significantly lower than the Compass laser. This substantially reduced the background noise on our CHIRPT data. This configuration was used to collect data displayed in all figures aside from those noted above. 

Since the exit pupil of high NA objective lenses often lies inside the objective case, access to the image plane conjugate to the object region (and modulator mask), where the horizontal slit was used to spatially filter the diffracted orders from the modulator (Fig.~2a), was not accessible in the initial CHIRPT microscope design. To overcome this limitation we designed the microscope with dual image relay systems to allow unrestricted access to the conjugate image plane. After spatial filtering and beam expansion, the beam was brought to a horizontal line focus on the modulation mask with a 150~mm focal length achromatic cylindrical lens (ThorLabs, ACY254-150-A-ML). The first image relay system consisted of a 125~mm focal length lens (ThorLabs, ACH-254-125-ML) after the modulator mask, followed downstream by a 100~mm focal length lens (ThorLabs, ACH-254-100-ML). The horizontal slit used to reject one of the diffracted orders was placed in the plane conjugate to the mask plane, which was located between these two lenses. The spatially filtered intensity profile was then image relayed to the object region with a 250~mm focal length tube lens (ThorLabs, ACH-254-250-ML) and an objective lens. For Fig.~3--Fig.~6, Fig.~8, Fig.~S\ref{fig:data_recovery}, and Fig.~S\ref{fig:aberration}, a 50$\times$/0.8~NA objective lens was used (Zeiss, N-Achroplan 50$\times$/0.8~NA Pol), and the overall magnification of the system from the modulator mask to the object plane was 95$\times$.

\subsection*{Modulation Mask} 
\begin{figure}[ht]
\begin{center}
\resizebox{3.5in}{!}{\includegraphics{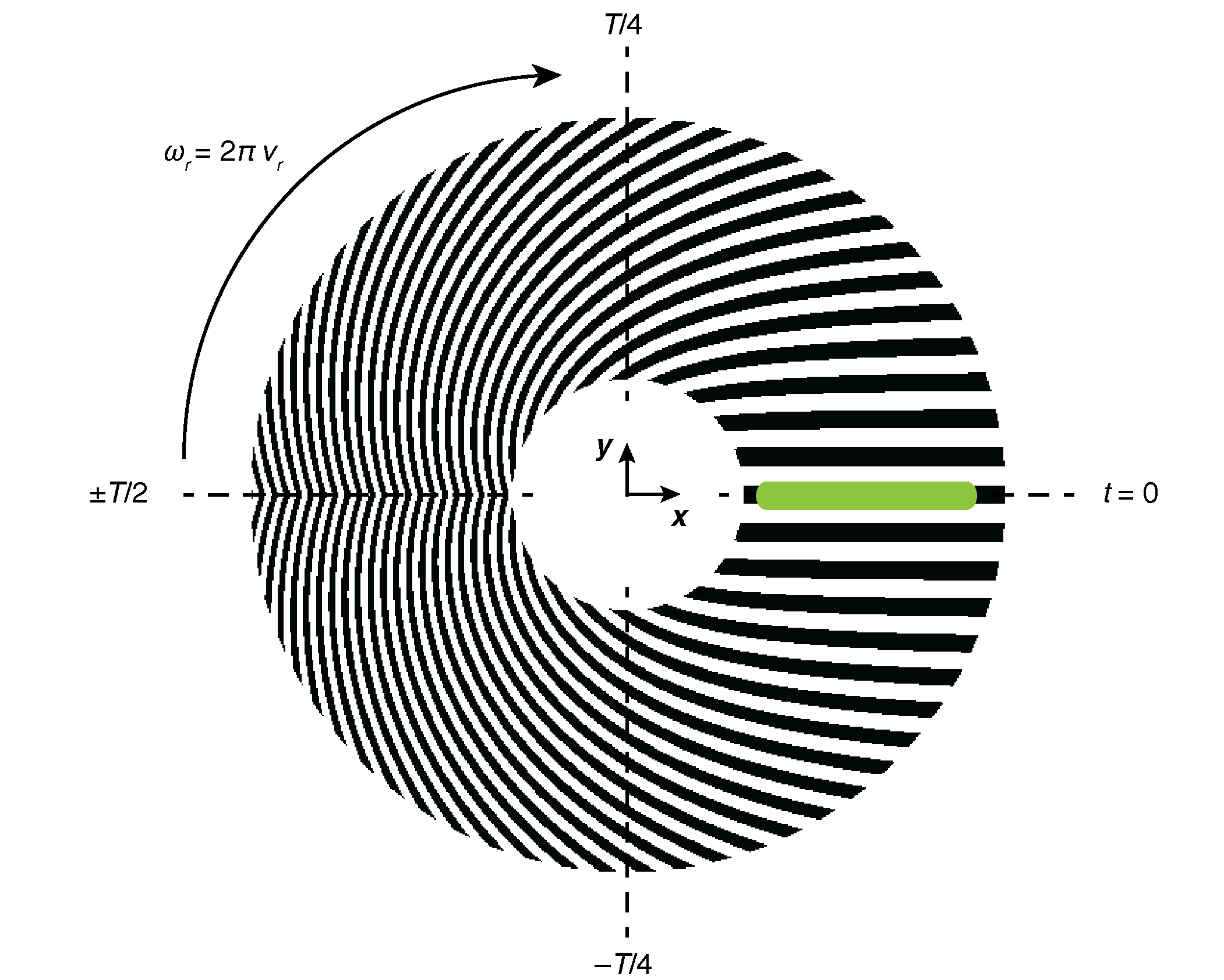}}
\caption{\label{fig:mask} A CHIRPT modulation mask with $\Delta k = 2/\mathrm{mm}$. As the mask is spun at constant angular velocity $\omega_r = 2 \pi \, \nu_r$, the modulation frequency varies linearly as a function of radial coordinate, $R$. Since the line focus is aligned along the $x$ axis, the modulation frequency experienced by the illumination light sheet is modulated with a temporal frequency that varies linearly with lateral position. Temporal modulation frequency, $\nu_t$, and lateral position, $x$, are related by the density of the mask, $\Delta k$, and the rotational frequency, $\nu_r$, by the expression: $\nu_t = \kappa_1 \, x = \Delta k \, \nu_r \, x$, where $\kappa_1 = \Delta k \, \nu_r$ is known as the chirp parameter. The inner and outer radii of the modulation pattern are 9.5~mm and 31~mm respectively. The green rectangle on the right side of the mask indicates the location of the laser line focus incident on the modulation pattern.}
\end{center}
\end{figure}

Modulation masks were generated with a rotationally-varied groove density according to the relation: $m(R,\varphi) = 1/2 \left\{ 1+\mathrm{sgn}[\cos(\Delta k \,  R \, \varphi)]\right\}$, where $(R,\varphi)$ represent polar coordinates relative to the center of the modulation pattern. A rendering of the mask is shown in Fig.~S\ref{fig:mask}. The modulation pattern was printed in aluminum onto a glass substrate (Projection Technologies, Dallas, Texas) at 3600 DPI. A center hole was cut with a water jet so the disk could be mounted on a custom chuck, and attached to a motor (Faulhaber, 2057S012BK1155). The desired angular velocity of the motor was set by the user in the data acquisition software and controlled by an external speed controller module (Faulhaber, MCBL3006).

\begin{figure}[ht]
\begin{center}
\resizebox{6.5in}{!}{\includegraphics{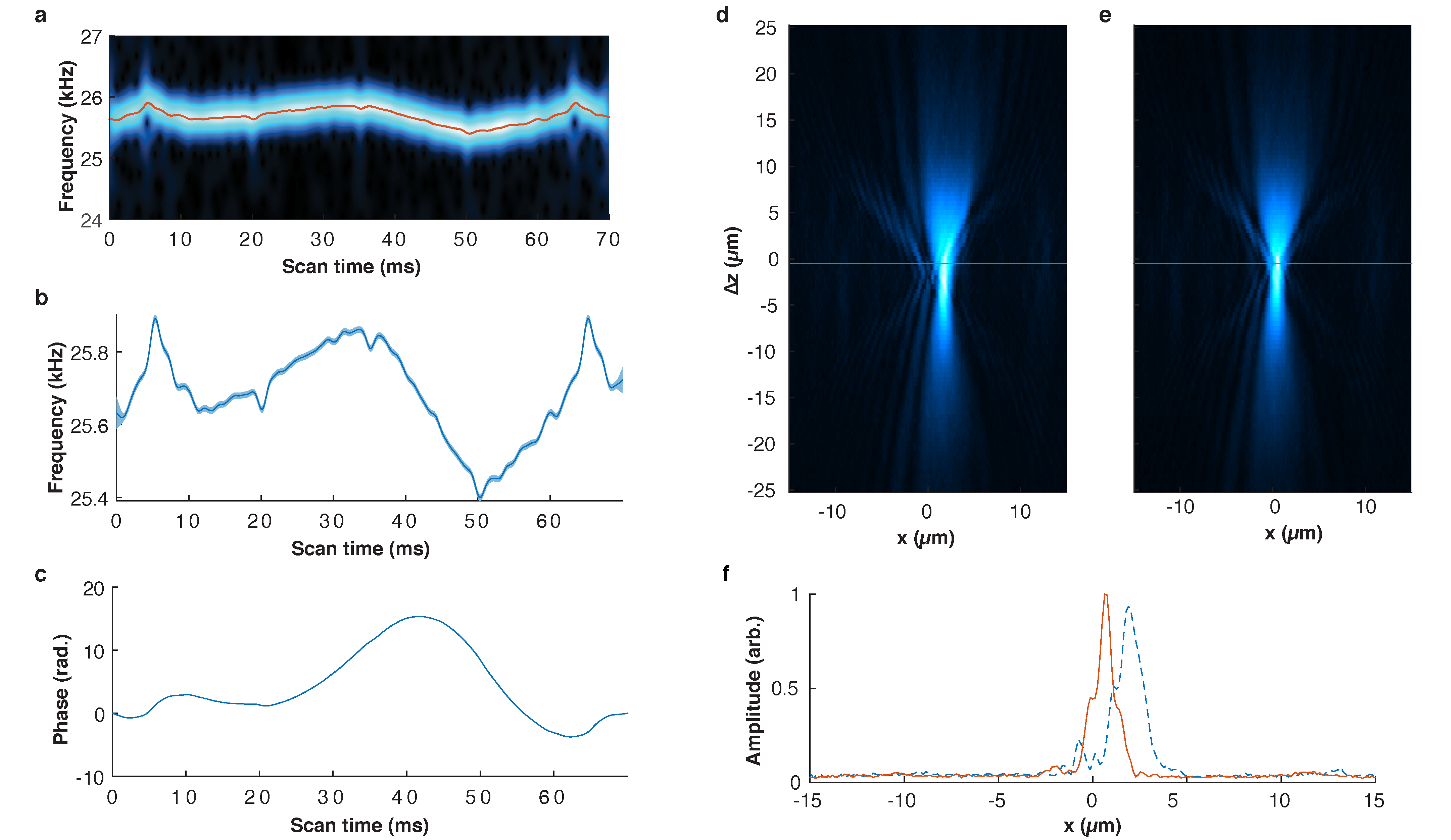}}
\caption{\label{fig:disk_phase} Disk phase aberration correction. (a) A spectrogram of the modulation frequency of the laser intensity at a single lateral location as a function of scan time. The computed centroid of the modulation frequency is shown by the overlaid red line. (b) The mean modulation frequency (solid line) and standard deviation (shaded region) were computed from 512 temporal scans. (c) The phase imparted by the modulation mask was uncovered by numeric integration of the mean modulation frequency. (d) A measured $(x,z)$ image of a 100~nm FND without correcting for the disk aberration phase. (e) A corrected image, computed by applying the conjugate disk aberration phase to the data. (f) The intensity of the corrected (solid line) and uncorrected (dashed line) fluorescence intensity near the focal plane.}
\end{center}
\end{figure}

\begin{figure}[!ht]
\begin{center}
\resizebox{6in}{!}{\includegraphics{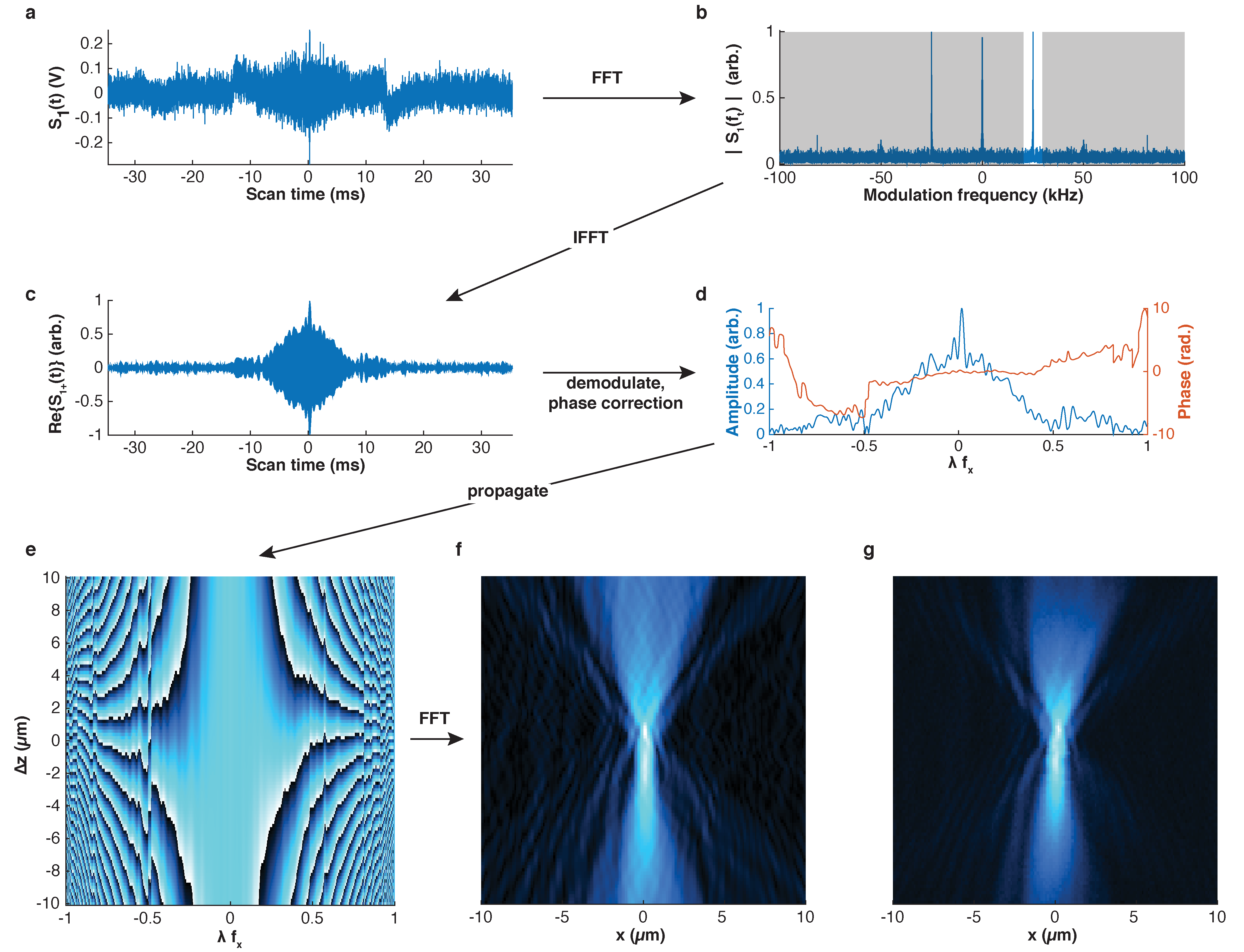}}
\caption{\label{fig:data_recovery} CHIRPT image reconstruction. (a) A single temporal scan collected by measuring fluorescent light from a fluorescent nanodiamond (FND) near the focal plane of the microscope with a 50$\times$/0.8 NA objective lens. (b) The spectral density of the measured temporal trace. (c) An inverse FFT (IFFT) is used to recover the complex CHIRPT trace in the temporal domain, $S_{1+}(t)$. (d) The CHIRPT trace is demodulated by the carrier frequency to remove the linear phase. Other phase disturbances, i.e., systematic aberration phase, or pupil phase, and the disk aberration phase are also removed to compensate for known aberrations to the system. (e) The demodulated and phase-corrected CHIRPT data is digitally propagated by applying the Ewald phase. (f) A one-dimensional FFT with respect to the lateral spatial frequency domain reveals a numerically-reconstructed two-dimensional image in $(x,z)$ space. (g) An image in the $(x,z)$ plane formed by physically scanning the FND along the defocus dimension.}
\end{center}
\end{figure}

To measure and correct for small modulation frequency errors arising from modulation mask centering misalignment, which inevitably occurs during mounting of the modulation disk, the illumination laser was brought to a focal point on the modulation mask with a second cylindrical lens with a focal length of 125 mm (Thorlabs, ACY254-125-A-ML) that was oriented perpendicularly to the permanent cylindrical lens. The throughput laser intensity as a function of scan time was measured with a photodiode for 512 traces. For each trace, a spectrogram was computed and the central modulation frequency was obtained by fitting the centroid of the spectrogram (Fig.~S\ref{fig:disk_phase}). The average modulation frequency was then computed from the set of 512 centroid frequency traces. If the modulation mask was mounted perfectly, the modulation frequency would be constant as a function of scan time since the focal spot samples the same position in space. Deviations from constant frequency correspond to imperfect mounting of the mask on the rotation axis of the motor. Mathematically, we can treat these deviations as a time dependent phase shift, such that the signal measured with the photodiode has the form:
\begin{equation}
V_m(t)  \propto  \cos \left[ 2 \pi \, \kappa_1 \, t \, x_s + \phi_m(t) \right] = \cos \left[ 2 \pi \, \nu_s \, t + \phi_m(t) \right]
\end{equation}	  
where $\kappa_1$ is the chirp parameter of the mask, $x_s$ is the lateral location of the focal spot on the mask, and $\phi_m(t)$ is the phase that causes deviations from a constant modulation frequency, $\nu_s$. This phase can be recovered from the modulation frequency measurement by numerically integrating the centroid frequency vs. time data and multiplying by $2 \pi$ (cf. Eq.~(2) of the manuscript). The conjugate of the mask phase was applied to measured CHIRPT data to correct for the phase disturbance.

All data presented in this work used mask designs with $\Delta k = 70/\mathrm{mm}$. For high-speed data in Fig.~S\ref{fig:fast_scanning}, a disk with four masks was utilized to increase imaging speeds.

\subsection*{Detection System} 
Fluorescent light intensity was measured with a photomultiplier tube (PMT). Data in Fig.~S\ref{fig:fast_scanning} were recorded with a Hamamatsu H9305-03. All other fluorescent images were measured with a Hamamatsu H7422P-40. 

For the high-speed data in Fig.~S\ref{fig:fast_scanning}, the signal from the PMT was high-pass filtered by a current preamplifier (Stanford Research Systems, SR570) with 10 kHz corner frequency and 12 dB/octave rolloff. Data in Fig.~3--Fig.~6, Fig.~8, Fig.~S\ref{fig:data_recovery}, and Fig.~S\ref{fig:aberration} were collected with a low-noise current amplifier (DHPCA-100, Femto Messtechnik GmbH, Berlin, Germany). Transmitted laser intensity presented in Fig.~7 was measured with a large-area photodiode (ThorLabs, DET100A). Amplified and filtered signals were digitized with a PCI data acquisition card (National Instruments, PCI-6110 with BNC-2090A breakout box). Data was displayed to the user and saved for post-processing with a custom C\# Windows application.

\subsection*{Fluorescent Objects} 
A 100 nm diameter fluorescent nanodiamond (FND) was used as a fluorescent probe for data presented in Fig.~3, Fig.~4, Fig.~8, Fig.~S\ref{fig:data_recovery}, and Fig.~S\ref{fig:aberration}. A 10 $\mu$L drop of FND slurry (ND-400NV-100nm-10mL, Ad\'{a}mas Nanotechnologies, Raleigh, NC) was drop cast onto a standard microscope slide and allowed to dry at room temperature. A cover slip was then placed over the sample to minimize spherical aberration.
	
A single 15-$\mu$m-diameter, shell-stained fluorescent microsphere in a prepared slide (LifeTechnologies, FocalCheck$^{\mathrm{TM}}$ Slide 1, Well A1) was used for CHIRPT data presented in Fig.~5. Although the slide contains fluorescent microspheres with a variety of fluorophores, the orange fluorescent spheres were most visible with the illumination wavelength of 532 nm.

The murine intestinal images presented in Fig.~6 were collected from a commercial prepared slide (LifeTechnologies, FluoCells\circledR Prepared Slide \#4). 

\subsection*{CHIRPT Reconstruction} 
Data processing was performed in MATLAB (The MathWorks, Natick, NJ). Image recovery for a single temporal data trace was achieved by a fast Fourier transform (FFT) operation. The lateral spatial axis was calibrated according to the relation $f_x \, x = \nu_t \, t$, which yields:
\begin{equation}
x = \frac{n_1}{n_2} \, \frac{\nu_t}{M \, \kappa_1} = \frac{n_1}{n_2} \,  \frac{\nu_t}{M \, \Delta k \, \nu_r} = \frac{n_1}{n_2} \, \frac{\nu_t}{\kappa_2}
\end{equation}
where $\nu_t$ is the temporal modulation frequency, $\nu_r$ is the rotation frequency of the mask, $\Delta k$ is a known design parameter of the mask, and $M$ is the magnification of the illumination microscope. The chirp parameter in the object region, $\kappa_2 = M \, \kappa_1$, was determined empirically by collecting CHIRPT signals while translating an isolated fluorescent object laterally, i.e., in the $x$ direction. The fluorescent object was typically a 15~$\mu$m shell-stained bead. The centroid modulation frequency of the fluorescence distribution was computed for each CHIRPT trace, which corresponded to a known lateral location. The chirp parameter was determined from the slope of a linear fit to the centroid frequency vs. lateral position data.

Refocusing of fluorescent light intensity was performed by applying a propagator in the lateral spatial frequency domain, which is directly proportional to time. The reconstruction processes is summarized visually in Fig.~S\ref{fig:data_recovery}. Propagation was applied to the positive frequency complex sideband of the measured CHIRPT time trace. This sideband was selected by filtering out only the positive temporal Fourier sideband that contains the modulation frequencies. The centroid of the positive frequency sideband was computed and used to demodulate the temporally-modulated complex data. The demodulated data were numerically propagated with a propagation phase conjugate to the axial phase component in Eq. (1) This phase varies linearly with defocus distance, $\Delta z$. After application of the Ewald propagation phase, a 1D FFT with respect to lateral spatial frequency then gives the propagated image. Mathematically, propagation can be described as: 
\begin{equation}
S_{1+}^{(r)}(x,z) = \hat{\mathcal{F}}_{f_x} \left\{ S_{1+}(f_{x,2}) \exp \left[ \mp \mathrm{i} \, 2 \pi \, \frac{n_2}{\lambda}\,  \Delta z  \left( \sqrt{1 - \left(\frac{\lambda}{n_1}\, f_{x,2} \right)^2} - 1\right) \right] \right\} 
\end{equation}
where the superscript (r) is used to denote the refocused data, $\hat{\mathcal{F}}_{f_x}$ is the Fourier transform operator with respect to lateral spatial frequency, and $S_{1+}(f_x)$ denotes the complex CHIRPT data, which is synthesized from the measured real-valued data in the temporal domain, $S_1(t)$, by a Hilbert transform. The sign of the argument used to propagate depended on which diffracted order was passed through the spatial filter (i.e., $j = \pm1$).
	
Data presented in Fig.~S\ref{fig:fast_scanning} was further processed by multitaper power spectral density (PSD) analysis using the function pmtm() in MATLAB with a time-bandwidth product of 5/2. The processed image was computed by taking the square-root of the PSD estimate.

\begin{table}[ht]
\begin{center}
\caption{\bf Data Acquisition Parameters}
\begin{tabular}{lccccc}
\hline
Figure & $\nu_r$ (s$^{-1}$) & $N_\mathrm{scans}$ & $N_\mathrm{pix}$ & $t_\mathrm{total}$ (s) & $\mathcal{D}$ \\
\hline
3 & 17 & 20 & 101 & 119 & 2D \\
4 & 17 & 25 & 101 & 148.9 & 2D \\
5 & 17 & 15 & 101 & 89.3 & 3D \\
6 & 17 & 15 & 101 & 443.7 & 3D \\
7 & 31 & 1 & 100 & 3.23 & 2D \\
8a & 14.3 & 1 & 1 & 0.07 & 2D \\
8b & 14.3 & 20 & 101 & 141.4 & 2D \\
S4 & 14.3 & 20 & 101 & 141.4 & 2D \\
S5 & 403 & 1 & 401 & 0.995 & 3D \\
\hline
\end{tabular}
  \label{tab:times}
\end{center}
\end{table}

Most data displayed in the main text were formed by averaging multiple images. Table~\ref{tab:times} summarizes the acquisition parameters for each image in this work, where $\nu_r$ is the frame rate of 1D data acquisition (2D after CHIRPT reconstruction), $N_\mathrm{scans}$ is the number of 1D data scans averaged for each position that the object was translated, $N_\mathrm{pix}$ is the number of positions at which data was collected, $t_\mathrm{tot}$ is the total image acquisition time, and $\mathcal{D}$ is the dimensionality of the reconstructed CHIRPT data. Note that the same data is displayed in Fig.~8 and Fig.~S\ref{fig:data_recovery}.

\subsection*{Confocal Imaging} 
Spinning-disk confocal images of murine intestinal tissue were collected with a commercial microscope base (IX81, Olympus) equipped with a spinning disk confocal head (CSU22, Yokogawa Instruments). Fluorescence was excited with an illuminating laser with 561 nm wavelength. An objective lens with NA of 0.5 focused the illuminating light into the specimen (Olympus UPlanFL N 20$\times$/0.5 Ph1). Fluorescent light was isolated from the illumination laser with an interference filter that had a 73 nm bandwidth centered at 617 nm  (FF02-617/73, Semrock).  Fluorescent images were formed on an electron multiplying charge coupled device (EMCCD) camera (Cascade II:1024, Photometrics). Image acquisition was controlled with SlideBook v.6 software (Intelligent Imaging Innovations). 

\section*{Theoretical Analysis of CHIRPT}
In CHIRPT, the spatial phase of coherent illumination beams is imprinted onto a temporal intensity modulation pattern that is unique to each $(x,z)$ position.  Here we compute the illumination intensity pattern by assuming spatially-coherent, monochromatic plane waves interfering in the object region. We then show that the spatial phase difference between the two interfering illumination beams is encoded as temporal modulations on the illumination intensity pattern.

CHIRPT encodes complex spatial image information into a temporal intensity pattern by measuring the radiant flux from the contrast distribution on a single-element photodetector as the illumination intensity pattern is modulated over time.  The CHIRPT signal has the form
\begin{equation}
S(t) = \int_{-\infty}^{\infty} \! \! \mathrm{d}^3 {\bf r} \, \, I_\mathrm{ill}({\bf r};t) \, C({\bf r}, t)
\end{equation}
where $I_\mathrm{ill}({\bf r};t)$ is the illumination pattern in the object region, and $C({\bf r}, t)$ is the contrast distribution.  The contrast distribution describes the method by which illumination intensity is transferred to contrast intensity.  For example, the contrast distribution can simply represent the transmittance of the object if the measured contrast intensity is the illumination source after propagation through an object displaying absorption.  In that case, $C({\bf r},t) = \mathbb{T}({\bf r},t)=1-\mathbb{A}({\bf r},t) $, where $\mathbb{T}$ is the intensity transmission of the object, and $\mathbb{A}$ is the intensity absorption of the object, and an image is formed from the object transmission.  The contrast can also represent the spatial concentration distribution of fluorophores in an object, scaled by the absorption cross section for the excitation light and efficiency of fluorescent emission.  Throughout this analysis, we shall assume that the contrast function is constant for the duration of a scan, so that $C({\bf r}, t) \rightarrow C({\bf r})$.  Moreover, we will assume that the contrast function is linearly proportional to the illumination intensity.

To compute the form of the illumination intensity, which arises from the spatial interference of the undiffracted and diffracted beams, consider two plane waves interfering in the lateral-axial plane, $(x,z)$, as shown in the object region in Fig.~1a.  We define the coordinate system such that the undiffracted beam (green) lies along the optic axis, ${\bf \hat{z}}$. Throughout the following analysis, we use a subscript pair $(j,s)$ to denote the diffracted order $j$ in region $s$, where $s = 1$ corresponds to the mask region and $s = 2$ corresponds to the object region. For example, the electric field of the undiffracted beam ($j = 0$) in the object region is:
\begin{equation}
v_{0,2}(x,z;t) = a_0 \, \ee^{\ii \, \bfk_{0,2} \cdot \bfr} = a_0 \, \ee^{\ii \, k_2 \, z}
\end{equation}
where $a_0$ is the amplitude of the field, $\bfk_{0,2}$ is the wave-vector of the undiffracted beam in the object region, $k_2 = \left| \bfk_2 \right| = 2\pi \, n_2/\lambda$ is the wavenumber, and $n_2$ is the refractive index in the object region.

Conversely, the first-order diffracted beam ($j = +1$) propagates at an angle with respect to the optic axis that varies with scan time, $\theta_2(t)$, so the wave-vector for the positive diffracted beam is:
\begin{equation}
\bfk_{1,2}(t) = k_2 \, \sin \theta_2(t) \, {\bf \hat{x}} + k_2 \, \sqrt{1 - \sin^2 \theta_2(t)} \, {\bf \hat{z}}.
\end{equation}
Additionally, the diffracted beam accumulates an optical path length (spatial phase) that varies with propagation angle, and corresponds to the pupil phase of the imaging system, $\Phi_\mathrm{pupil}(f_x)$.  Since the lateral spatial frequency, $f_x$, is directly proportional to scan time, $t$, the electric field of the diffracted beam in the object region is:
\begin{eqnarray}
v_{1,2}(x,z;t) & = & a_1 \, \ee^{\ii \, \bfk_{1,2} \cdot \bfr} \nonumber \\
 & = &  a_1 \, \exp \left[ \ii \, \Phi_\mathrm{pupil}(t) \right] \, \exp \left[\ii \, k_2 \, x \, \sin \theta_2(t) \right] \,  \exp \left[ \ii \, k_2 \, z \, \sqrt{1 - \sin^2 \theta_2(t)} \right] \, 
\end{eqnarray}

From Fig.~1a, it is apparent that the propagation angle in the object region, $\theta_2(t)$, is related to the angle of diffraction from the mask, $\theta_1(t)$.  Specifically, these angles are related by the magnification of the image relay system, $M$, which is the ratio of the tube lens focal length to the objective lens focal length, i.e., $M = F_t/F_o$.  Moreover, the diffraction angle in the mask region is related to the spatial frequency of the modulation pattern on the mask according to the expression: $n_1 \, \sin \theta_1(t) = \lambda \, f_{x,1}(t)$, where $f_{x,1}(t)$ is the spatial frequency of the mask, and $n_1$ is the refractive index of the medium that the mask is immersed in.  Altogether:
\begin{equation}
\sin \theta_2(t) = \frac{F_t}{F_o} \, \sin \theta_1(t) = \frac{\lambda}{n_1} \, M \, f_{x,1}(t).
\label{sin_angle_object}
\end{equation}  

To derive the time-dependent spatial frequency, we consider the modulation mask.  In polar coordinates, the modulation mask has the form:
\begin{equation}
m(R,\varphi) = \frac{1}{2} + \frac{1}{2} \mathrm{sgn} \left[ \cos \left( \Delta k \, R \, \varphi \right) \right]
\end{equation}
where $\Delta k$ describes the density of features on the mask and has units of spatial frequency, and the function $\mathrm{sgn}[\cdot]$ is included to account for the binary amplitude modulation provided by the modulation mask. This expression represents the full two-dimensional pattern of the mask, while we are interested in the region sampled by the line focus, located at $y=0$, and approximately centered on one side of the mask such that the illumination distribution in $x$ lies between the inner and outer radii of the mask (Fig.~S\ref{fig:mask}).  To account for the changing modulation pattern sampled by the line focus during mask rotation, and thus find the local spatial frequencies as a function of scan time, we consider the mask:
\begin{equation}
m(R,\varphi - \varphi_0) = \frac{1}{2} + \frac{1}{2} \mathrm{sgn} \! \left\{ \cos \! \left[\Delta k \, R \, \left(\varphi - \varphi_0 \right) \right] \right\} \end{equation}
where $\varphi_0$ describes the rotation angle of the mask.  Since the mask rotates at a constant angular velocity $\omega_r = 2 \pi \, \nu_r$, where $\nu_r$ is the rotational frequency of the mask,  the rotation angle is: $\varphi_0 = 2 \pi \, \nu_r \, t$.

To uniquely determine the spatial phase of the illumination microscope, only the first diffracted order from the mask is permitted to interfere with the undiffracted beam in the object plane.  Higher diffracted orders resulting from the binary modulation scheme are omitted in the following analysis, and also excluded in our CHIRPT experimental design, so that the mask we consider is:
\begin{equation}
m(R,\varphi - \varphi_0) = \frac{1}{2} + \frac{1}{2} \cos \! \left[\Delta k \, R \, \left(\varphi - \varphi_0 \right) \right] = \frac{1}{2} + \frac{1}{2} \cos \! \left[\, \phi \! \left( R, \varphi_0 \right) \, \right]
\end{equation}

The linear spatial frequency on the mask in the lateral dimension is \cite{Goodman:2005jt}:
\begin{equation}
f_{x,1}(t)  =  \frac{1}{2 \pi} \, \frac{\partial \phi ({\bf r}, \varphi(t))}{ \partial x}\\
\end{equation}
The local region sampled by the beam is along the line where $\varphi = 0$, so the local linear spatial frequency imparted to the beam is:
\begin{equation}
f_{x,1}(t) = \frac{1}{2 \pi} \left[ \cos \! \varphi \, \frac{\partial \phi({\bf r}, \varphi)}{\partial r}  - \frac{1}{r} \, \sin \! \varphi \, \frac{\partial \phi({\bf r}, \varphi)}{\partial \, \varphi}  \right]_{\varphi = 0} = \kappa_1 \, t
\end{equation}
where $\kappa_1 \equiv \Delta k \, \nu_r$ is the so-called chirp parameter, which describes the linear increase in modulation frequency with lateral position on the mask. Equation~\eqref{sin_angle_object} becomes:
\begin{equation}
\sin \theta_2(t) = \frac{\lambda}{n_1} \, M \, \Delta k \, \nu_r \, t = \frac{\lambda}{n_1} \, \left( M \, \kappa_1 \, t \right).
\end{equation}
The lateral spatial frequency in the object region is then:
\begin{equation}
f_{x,2}(t) = \frac{n_2 \, \sin \theta_2(t)}{\lambda} = \frac{n_2}{n_1} \, M \, \kappa_1 \, t
\end{equation}

The electric field of the diffracted beam can now be written:
\begin{eqnarray}
v_{1,2}(x,z;t) & = & a_1 \, \exp \left[\ii \, 2 \pi \, \left(\frac{n_2}{n_1} \, M \, \kappa_1 \, t\right) \, x \right]  \exp \left[\ii \, 2 \pi \, \frac{n_2}{\lambda} \, z \, \sqrt{1 - \left( \frac{\lambda}{n_1} \, M \, \kappa_1 \,t \right)^2}\right]  \ee^{\ii \, \Phi_\mathrm{pupil}(t)} \nonumber \\
	& = &  a_1 \, \exp[\ii \, \phi_\mathrm{lateral}(x;t)] \, \exp[\ii \, \phi_\mathrm{axial}(z;t)]  \,  \exp[\ii \, \Phi_\mathrm{pupil}(t)]
\end{eqnarray}
Here we see that the diffracted order accumulates a linear spatial phase that varies with lateral position, a circular spatial phase that varies with axial position, and a pupil phase that varies only with scan time.

The axial phase component describes the optical path length (OPL) change imparted to the diffracted beam as it is scanned across the pupil as a function of time. The magnitude of the linear spatial frequency in the object region is $f_2 = k_2/(2 \pi) = n_2/\lambda$, and the axial phase can be written as:
\begin{equation}
\phi_\mathrm{axial}(z;t) =  k_2 \, z \, \sqrt{1- \left( \frac{\lambda}{n_1} \, M \, \kappa_1 \, t \right)^2} =  2 \pi \, f_2 \, z \, \sqrt{1- \left[\frac{f_{x,2}(t)}{f_2} \right]^2}
\end{equation} 
This phase corresponds to physical defocus in the imaging system, which is represented schematically by the curved orange wavefronts in Fig.~1a, and is consistent with the angular spectrum representation \cite{L:2012mz}. This phase is formally equivalent to the Ewald phase. 

We note that it is possible to construct an imaging system for which the paraxial approximation applies. Applying a small-angle approximation to this expression, wherein we assume $\sin^2 \theta_2(t) = \left[ \left(\lambda \, M \, \kappa_1 \, t \right)/n_1 \right]^2 \ll 1$, or equivalently, $ \left [f_{x,2}(t)/f_2 \right]^2 \ll 1$, the axial phase can be written as a quadratic phase variation:
\begin{equation}
\phi_\mathrm{axial}(x,z;t)  \approx 2 \pi \, \frac{n_2}{\lambda} \, z \, \left\{1 - \frac{1}{2} \left[\frac{\lambda}{n_2} \,f_{x,2}(t) \right]^2 \right\}
\end{equation}
This phase variation is equivalent to the Fresnel propagation phase \cite{Goodman:2005jt}, and corresponds to physical defocus in the imaging system, i.e., axial displacement away from the object focal plane. 

Finally, the illumination intensity in the object region is computed by a coherent summation of the diffracted and undiffracted beams: $I_\mathrm{ill}(x,z;t) = \left| v_{0,2}(x,z;t) + v_{1,2}(x,z;t) \right|^2$.  We assume beam amplitudes $a_0 = 1/2$ and $a_1 = 1/\pi$ based on the diffraction efficiency from a square grating. The illumination intensity evaluates to:
\begin{eqnarray}
I_\mathrm{ill}(x,z;t) & = & I_0 + I_1(x,z;t) = \left(\frac{1}{4} + \frac{1}{\pi^2} \right) \nonumber \\
	& & + \frac{1}{\pi} \, \cos \! \left\{ 2 \pi \, \left( \frac{n_2}{n_1} M \,\kappa_1 \, t \right)  \, x  + \frac{2 \pi}{\lambda} \, n_2 \, z \, \left[\sqrt{1 - \left( \frac{\lambda \, M \, \kappa_1 \, t}{n_1} \right)^2} - 1 \right]  +  \Delta \Phi_\mathrm{pupil}(t) \right\}.
\label{illumination_general}
\end{eqnarray}
The illumination intensity consists of a constant (DC) term, $I_0$, and a temporally varying (AC) term, $I_1(x,z;t)$.  The CHIRPT image is encoded by the AC component of the illumination intensity:
\begin{equation}
S_1(t) = \iint_{-\infty}^{\infty} \! \! \mathrm{d}x \, \mathrm{d}z \, \, I_1(x,z;t) \, C(x,z)
\label{signal_equation}
\end{equation}

The argument of the cosine term in Eq.~\eqref{illumination_general} contains a two-dimensional map of the spatial phase variation in the object region due to system defocus as well as the pupil phase of the imaging system.

\subsection*{Digital Aberration Correction}
The differential phase between the two illumination beams encodes the pupil phase of the imaging system as well as physical defocus.  The pupil phase accounts for deviations in the optical path length due to systematic aberrations in the imaging system.  Since this phase difference is encoded in the temporal modulation frequency of the measured signal, it can by digitally removed to correct these aberrations.

\begin{figure}[!ht]
\begin{center}
\resizebox{6.75in}{!}{\includegraphics{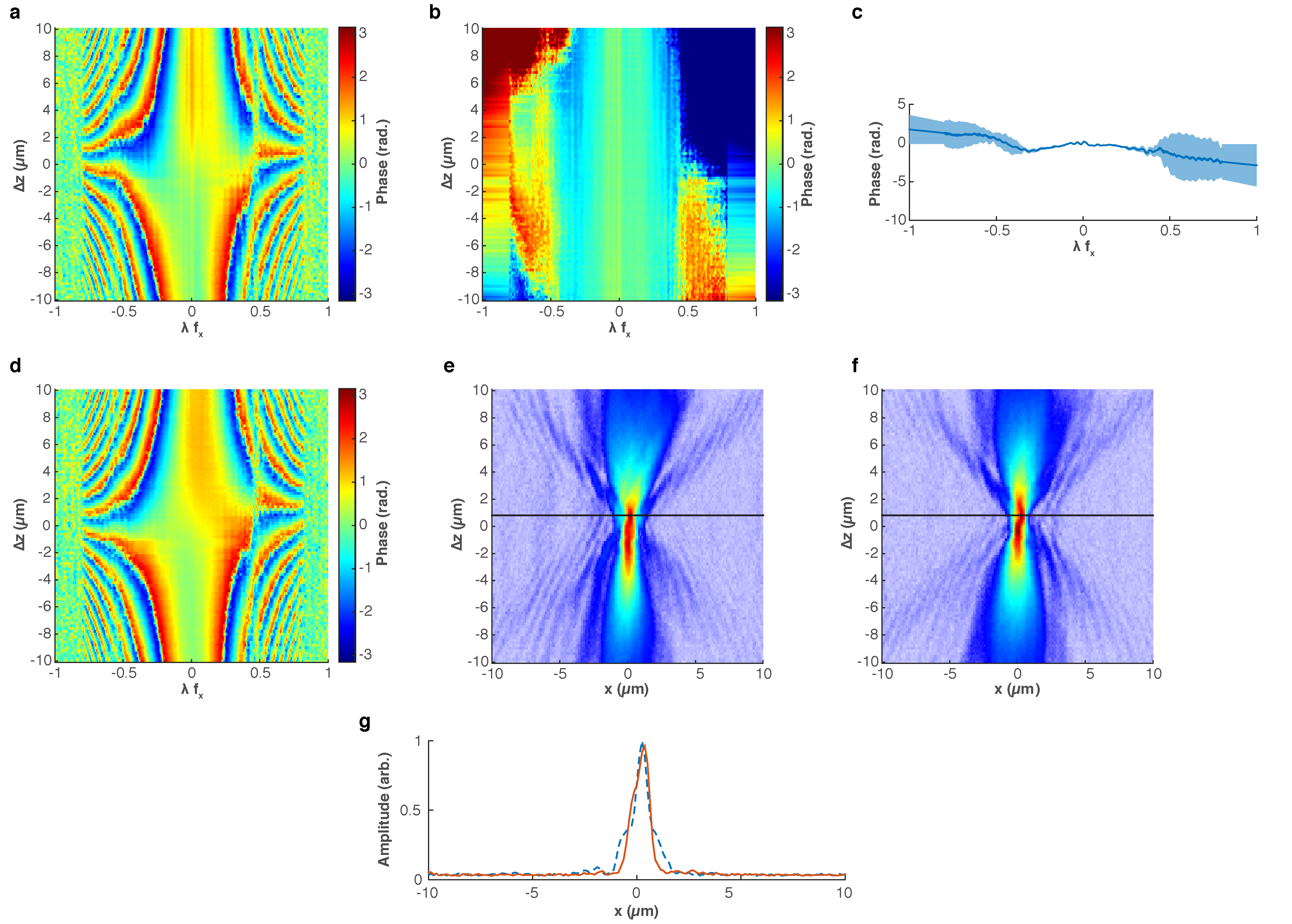}}
\caption{\label{fig:aberration} Systematic aberration phase correction. (a) Measured phase in the lateral spatial frequency domain, collected with a 95x/0.8~NA CHIRPT microscope by physically defocusing a 100~nm FND. This phase map is the average of 20 scans. (b) Removing the theoretical Ewald phase from the 2D phase map in (a) reveals the residual phase as a function of defocus. (c) The average (solid line) and standard deviation (shaded region) of the residual phase in (b) representing the systematic aberration phase of the imaging system. (d) A corrected phase map, formed by removing the mean aberration phase in (c) from the measured phase in (a). (e) An image of the FND before removing the systematic aberration phase, and, (f) after removing the aberration phase. (g) The intensity of the uncorrected image (dashed line) and corrected image (solid line) at the location indicated by the solid black lines in (e) and (f). Total collection time for all 20 scans was 141.4 s.}
\end{center}
\end{figure}

An example of systematic aberrations in an imaging system is shown in Fig.~S\ref{fig:aberration}, where the fluorescence intensity emitted by a 100~nm fluorescent nanodiamond (FND) was measured as a function of system defocus. The average of 20 phase measurements as a function of physical defocus was computed from the data. In the absence of pupil phase, the phase recovered from this measurement would be identical to the phase defined by the Ewald circle, i.e.:
\begin{equation}
\phi_{\mathrm{Ewald}}(f_{x,2}; \, z) = 2 \pi \, \frac{n_2}{\lambda} \, \Delta z \, \left[\sqrt{1 - \left( \frac{\lambda}{n_2} \, f_{x,2} \right)^2} - 1 \right]
\end{equation}
To uncover the pupil phase we computed the Ewald phase, with $\Delta z$ = 0 at the nominal focus of the measured image, and removed it from the measured phase. The resulting phase map showed a relatively constant phase over a range of defocus values.  The mean pupil phase was computed  and removed from the measured data. Two-dimensional images of the FND fluorescence intensity in $(x,z)$-space were computed from the measured data and the phase-corrected data. The symmetry of the intensity distribution vs. defocus was improved in the image when the residual pupil phase was removed.  Intensities of each image near the focal plane show a significant improvement of the focal quality in the phase-corrected image.

\subsection*{High-Speed CHIRPT with Fluorescent Molecules}
Since excited fluorescent molecules emit light spontaneously, there is a delay between excitation of a fluorophore and emission of a fluorescent photon.  This delay is characterized by the fluorescent lifetime, $\tau_f$, of the molecule. A requirement of CHIRPT imaging is that emitted fluorescent light must follow the temporally-modulated illumination intensity pattern.  This requirement is met when the modulation frequency of excitation is well below the inverse lifetime of the molecule, i.e., $\nu_t \ll \tau_f^{-1}$.  If the temporal frequency of excitation does not fulfill this requirement, the emitted fluorescence intensity acquires an appreciable temporal phase shift, $\phi_f$, as well as a change in the modulation depth, $m_f$.  The phase shift contributes to a phase delay in the measured signal, while loss of modulation depth reduces the signal-to-noise.

The emitted fluorescence intensity is directly proportional to the excitation intensity, $I_\mathrm{ex}(x,z;t)$, which is proportional to the product of the illumination intensity and the contrast distribution.  Taking fluorescent lifetime into account, the temporally-varying fluorescence intensity emitted by a fluorescent molecule becomes:
\begin{equation}
I_\mathrm{em}(x,z;t) \propto m_f \, \cos \left\{ \phi_f + 2 \pi \, \Delta x \, \left( \frac{n_2}{n_1} M \, \kappa_1 \, t \right) + k \, n_2 \,  z \left[\sqrt{1 - \left( \frac{\lambda \, M \, \kappa_1 \, t}{n_1} \right)^2} - 1 \right]   \right\}
\end{equation}
where we have retained only the the portion of the fluorescence intensity that encodes the CHIRPT image -- the temporally varying signal.

Fluorescent lifetime imaging with temporally-modulated excitation sources has been studied elsewhere \cite{Lakowicz:1982rc}.  Following this analysis, and consistent with the theoretical analysis of CHIRPT imaging presented above, we consider fluorophores excited by a sinusoidally-varying intensity pattern.  Assuming single-exponential lifetime decays, the phase delay of the emitted fluorescence intensity in CHIRPT imaging evaluates to:
\begin{equation}
\phi_f = 2 \pi \, \kappa_1 \, x \, \tau_f,
\end{equation}
and the depth of modulation is: 
\begin{equation}
m_f = \sqrt{1+\left(2 \pi \, \kappa_1 \, x \, \tau_f \right)^2}.
\end{equation}
Thus we can write the temporally-varying signal from the photodetector [Eq.~\eqref{signal_equation}] as:
\begin{eqnarray}
S_1(t) & \propto & \iint \limits_{-\infty}^{\infty} \! \! \mathrm{d}x \, \mathrm{d}z \, \, \sqrt{1+\left(2 \pi \, M \, \kappa_1 \, x \, \tau_f \right)^2}  \nonumber \\
	& & \times \, \cos \left\{ 2 \pi \, \kappa_1 \, x \, \tau_f + 2 \pi \, \Delta x \, \left( \frac{n_2}{n_1} M \, \kappa_1 \, t \right)   + k \, n_2\, z \left[\sqrt{1 - \left( \frac{\lambda \, M \, \kappa_1 \, t}{n_1} \right)^2} - 1 \right]  \right\}
\end{eqnarray}

\begin{figure}[!ht]
\begin{center}
\resizebox{5in}{!}{\includegraphics{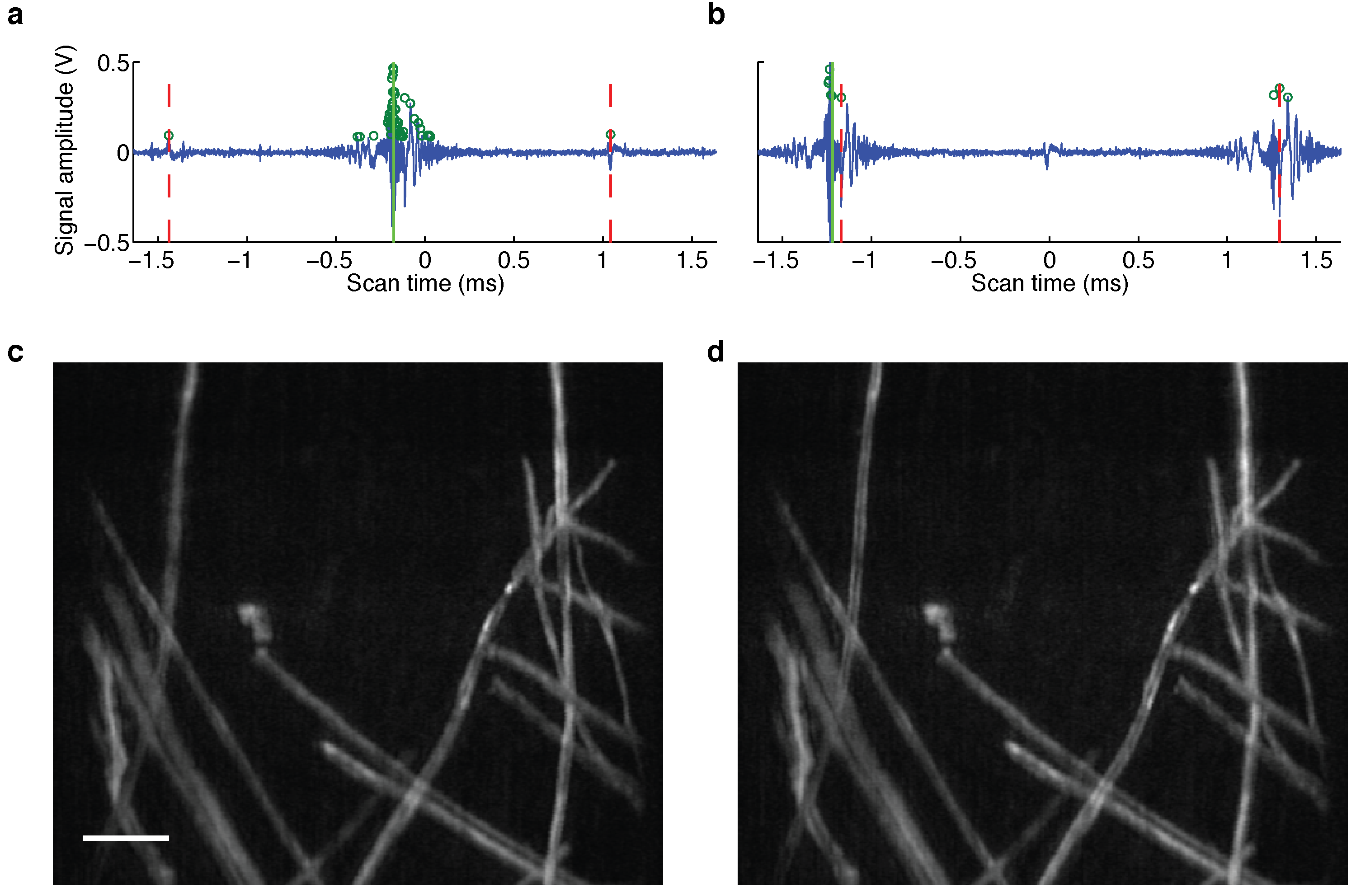}}
\caption{\label{fig:fast_scanning} High-speed epi-fluorescent CHIRPT images of lens tissue stained with highlighter ink. Scan times were measured by (a) using repetitive features in the temporal scan, and (b) adjusting the trigger delay of the DAQ card to collect two successive central peaks located near the center of the scan time, i.e., $t = 0$. Scan periods were found to be 2.48 ms in each case, corresponding to data acquisition for a full 2D $(x,z)$ plane at 403 frames/s. Each image is composed of 401 vertical scans for a total acquisition time of 995~ms for the full volume. In both (a) and (b) the temporal waveforms are the mean of 401 scans at varied vertical positions. (c) and (d), Digitally propagated images of the lens tissue, with the amplitude of each image estimated with a multi taper analysis (time-bandwidth product = 5/2). The propagation distances were -42~$\mu$m and +20~$\mu$m in panels (c) and (d) respectively. Scale bar: 100~$\mu$m.}
\end{center}
\end{figure}

The expression above indicates that for the recorded signal to reproduce the modulation frequency patten of the illumination intensity, the phase shift due to the fluorescent lifetime must be much less than unity: $2 \pi \, \kappa_1 \, x \, \tau_f \ll 1$.  Assuming a typical fluorescent lifetime of 10~ns \cite{Lakowicz:2010yq}, we computed this phase delay for data presented in Fig.~S\ref{fig:fast_scanning}, which had the highest temporal modulation frequency in this work, to be $\phi_f \approx 0.036$~radians.  

\subsection*{Vertical Diffraction from the CHIRPT Mask}
The CHIRPT illumination intensity pattern is formed by allowing only one of the diffracted beams to interfere with the undiffracted beam in the object region.  This is accomplished with a horizontally-oriented slit placed near the pupil plane of the illumination lens that passes only a portion of the undiffracted beam and a portion of one of the positive or negative diffracted orders (Fig.~2a).  This separation is possible due to a non-zero spatial frequency component in the vertical ($y$) direction that is independent of the rotation angle of the mask. Evaluating the local linear spatial frequency \cite{Goodman:2005jt} imparted by the modulator in the vertical dimension, we find:
\begin{equation}
f_{y,1} =  \frac{1}{2 \pi} \, \frac{\partial \phi ({\bf r}, \varphi)}{ \partial y} = \frac{1}{2 \pi} \left[ \sin \! \varphi \, \frac{\partial \phi({\bf r}, \varphi)}{\partial r}  + \frac{1}{r} \, \cos \! \varphi \, \frac{\partial \phi({\bf r}, \varphi)}{\partial \, \varphi}  \right]_{\varphi = 0}  =  \frac{\Delta k}{2 \pi}
\end{equation}
The non-zero spatial frequency in the vertical dimension causes the diffracted order to propagate at an angle with respect to the optic axis, $\psi_1$. In the pupil plane of the imaging system, the displacement of the diffracted beam from the optic axis is given by:
\begin{equation}
F_t \, \sin \psi_1 = F_t \,  \frac{\lambda \, f_{y,1}}{n_1} = F_t \, \frac{\lambda \, \Delta k}{2 \pi \, n_1}.
\end{equation}
where $F_t$ is the focal length of the tube lens. This displacement permits insertion of the horizontal slit to select only one diffracted order. 

For the data shown in Fig.~S\ref{fig:fast_scanning}, a modulation disk with 4 masks was used to increase image acquisition speed by providing a full temporal scan in only 90~degrees of mask rotation.  To combine multiple mask patterns onto a single disk, the pattern was phase-wrapped in angle.  Consequently, the density of features in the vertical dimension sampled by the incident line-focused beam increases as well.  In general, the spatial frequency in the vertical dimension increases by a factor of $N$ in this scenario, where $N$ is an integer number of masks per disk. Therefore the displacement from the optic axis in the pupil plane is increased by a factor of $N$ as well.

\subsection*{Depth of Field in CHIRPT}
One important advantage of CHIRPT imaging is an increased depth of field (DOF) as compared to conventional linear imaging.  In tightly-focused imaging techniques, such as laser-scanning microscopy, the depth of field and lateral spatial resolution are both set by the numerical aperture (NA) of the illumination objective, and the DOF varies inversely as the square of the NA.  CHIRPT does not suffer this same limitation because collimated beams with varied propagation angles are utilized to sequentially measure the spatial frequency content of an object.  Since the CHIRPT signal is generated by the temporally varying part of the illumination intensity, $I_1(x,z;t)$, the depth of focus is dictated by the region over which the intensity modulation is non-zero.  This region is formed by the interference of the re-imaged diffracted orders.

In the theoretical analysis thus far, we have assumed plane-wave illumination.  This causes both the field of view (FOV) and depth of field (DOF) to be infinite.  Let us assume instead that the illumination beam incident on the modulation mask has a Gaussian intensity profile in the lateral dimension and is collimated in the axial dimension.  We may write the incident illumination field propagating along the optic axis as
\begin{equation}
E_\mathrm{inc}(x,z) = u(x) \,  \mathrm{e}^{\mathrm{i} \, k_0 \, z} = \mathrm{e}^{-(x-x_0)^2/w^2} \,  \mathrm{e}^{\mathrm{i} \, k_0 \, z}
\end{equation}
where $w$ is the width of the Gaussian distribution. 

Due to diffraction from the modulation mask, the amplitude distribution propagates at an angle with respect to the optic axis in the object region (Fig.~1a).  We denote the amplitude of a diffracted order $j$ in the object region as $u_{j'}(x,z;t)$.  The amplitude of the undiffracted beam in the object region is simply given by the illumination amplitude scaled by the magnification of the illumination system:
\begin{equation}
u_{0}(x,z;t) = u_0(x/M) = \mathrm{e}^{-M^2 \, (x-x_0)^2/w^2}.
\end{equation}
The diffracted order, $j=1$, propagates at an angle with respect to the optic axis that depends on the rotational position of the mask and therefore depends on time [Eq.~\eqref{sin_angle_object}]. To simplify the following computations, we apply the approximation that upon diffraction from the mask the amplitude distribution simply rotates by the diffraction angle, i.e., there are no distortions to the amplitude distribution.  In the rotated coordinate basis $(x' , z')$, where $z'$ is the optic axis of the diffracted beam in the object region, the spatial distribution of the diffracted order is
\begin{equation}
u_1(x'/M) = \mathrm{e}^{-M^2 \, (x'-x'_0)^2/w^2}
\end{equation}

The DOF is dictated by the region of overlap of the diffracted and undiffracted spatial distributions:
\begin{equation}
\Lambda(x,z;t) = u_{0}(x,z;t) \, u_{1}(x,z;t) = u_{0}(x) \, u_{1}(x')
\end{equation}
The diffracted order distribution can be written in the unrotated coordinate basis, $(x,z)$, via the coordinate transformation:
\begin{equation}
x' = x \, \cos  \theta_2(t) - z \, \sin  \theta_2(t)
\end{equation}
Assigning the centroid of each distribution to the origin, $x_0=x_0'=0$, we can write:
\begin{equation}
\Lambda(x,z,t) = \exp \left[ -\frac{x^2}{w^2} \right] \, \exp \left[-\frac{x^2 \, \cos^2 \theta_2(t)}{w^2} - \frac{z^2 \, \sin^2 \theta_2(t)}{w^2}  + \frac{2 \, x \, z \, \sin \theta_2(t) \, \cos \theta_2(t)}{w^2} \right]
\label{dof_full}
\end{equation}
Using the shorthand notation $\Gamma \equiv \lambda \, M \, \kappa_1 \, t$, the overlap distribution in the object region is:
\begin{equation}
\Lambda(x,z,t) = \exp \left[ -\frac{x^2}{w^2} \right]  \, \exp \left[-\frac{\left(1 - \Gamma^2 \right) \, x^2}{w^2}  - \frac{\Gamma^2 \, z^2}{w^2} + \frac{2 \, \Gamma \, \sqrt{1- \Gamma^2} \, x \, z}{w^2} \right]
\label{dof_full_gamma}
\end{equation}

Although the DOF varies slightly with position in the object region, it is instructive to examine the DOF along the optic axis, i.e., $x=0$:
\begin{equation}
\Lambda(x=0,z,t) = \exp \left[ - \left(\frac{\lambda \, M \, \kappa_1 \, t}{w}\right)^2 \, z^2\right]  = \exp \left[ - \left(\frac{\lambda \, f_{x,2}(t)}{w}\right)^2 \, z^2\right]
\label{dof_on_axis}
\end{equation}
where $f_{x,2}(t)$ is the lateral spatial frequency in the object region.  We define the DOF along the optic axis, $\mathrm{DOF}_z$, as the full-width of the overlap distribution at the $1/e$-point, which we can extract from Eq.~\eqref{dof_on_axis}:
\begin{equation}
\mathrm{DOF}_z(t) = \frac{2 \, w}{\lambda \, M \, \kappa_1 \, t} = \frac{2 \, w}{\lambda \, f_{x,2}(t)}
\label{dof_z}
\end{equation}
Since the highest spatial frequency that contributes to the measurement corresponds to the maximal scan times, $t= \pm T/2$, Eq.~\eqref{dof_z} can be recast in terms of the numerical aperture of the imaging system to find the DOF for the full numerical aperture of the imaging system.
\begin{equation}
\mathrm{DOF}_z = \frac{2 \, w}{\lambda \, M \, \kappa_1 \, T/2} = \frac{4 \, w}{\lambda \, M \, \Delta k \, \nu_r \, T} = \frac{4 \, w}{\lambda \, M \, \Delta k} = \frac{2 \, w}{\mathrm{NA}}
\end{equation}
Unlike conventional light focusing, where the DOF varies as $1/\mathrm{NA}^{2}$, the DOF in CHIRPT at the full numerical aperture of the system varies as $1/\mathrm{NA}$.  This allows for high resolution imaging from a much broader depth.

\bibliography{refs,refs2}


\end{document}